\input epsf.tex
\input phyzzx

\font\textmi=cmmi9
\def\mathfig{\textfont1=\textmi 
\textfont0=\ninerm
}
\def\eatOne#1{}

\global\newcount\figno \global\figno=0

\def\ifundef#1{\expandafter\ifx%
\csname\expandafter\eatOne\string#1\endcsname\relax}

\def\mfig#1#2{\global\advance\figno by1%
\relax#1\the\figno%
\warnIfChanged#2{\the\figno}%
\edef#2{\the\figno}%
\reflabeL#2%
\ifWritingAuxFile\immediate\write\auxfile{\noexpand\xdef\noexpand#2{#2}}\fi%
}

\newif\ifFiguresInText\FiguresInTexttrue
\newwrite\capfile
\newwrite\figfile

\def\PlaceTextFigure#1#2#3#4{%
\vskip 0.5truein%
#3\hfil\epsfbox{#4}\hfil\break%
\hfil\vbox{\ninebf Figure #1\mathfig #2}\hfil%
\vskip10pt}
\def\PlaceEndFigure#1#2{%
\epsfysize=\vsize\epsfbox{#2}\hfil\break\vfill\centerline{Figure #1.}\eject}

\def\LoadFigure#1#2#3#4{%
\ifundef#1{\phantom{\mfig{}#1}}\fi
\ifWritingAuxFile\immediate\write\auxfile{\noexpand\xdef\noexpand#1{#1}}\fi%
\ifFiguresInText
\PlaceTextFigure{#1}{#2}{#3}{#4}%
\else
\if@FigureFileCreated\else%
\immediate\openout\capfile=\jobname.caps%
\immediate\openout\figfile=\jobname.figs%
\fi%
\immediate\write\capfile{\noexpand\item{Figure \noexpand#1.\ }#2.}%
\immediate\write\figfile{\noexpand\PlaceEndFigure\noexpand#1{\noexpand#4}}%
\fi}

\def\listfigs{\ifFiguresInText\else%
\vfill\eject\immediate\closeout\capfile
\immediate\closeout\figfile%
\centerline{{\bf Figures}}\bigskip\frenchspacing%
\input \jobname.caps\vfill\eject\nonfrenchspacing%
\input\jobname.figs\fi}

\newif\ifWritingAuxFile
\newwrite\auxfile
\def\SetUpAuxFile{%
\xdef\auxfileName{\jobname.aux}%
\inputAuxIfPresent{\auxfileName}%
\WritingAuxFiletrue%
\immediate\openout\auxfile=\auxfileName}


\hsize=6.5truein 
\voffset=0.1truein
\vsize=8.51truein
\def\TITLEPAGE{\frontpagetrue}
\def\PUPT#1{\hbox to\hsize{\tenpoint \baselineskip=12pt
        \hfil\vtop{
        \hbox{\strut PUPT-96-#1}
}}}
\def\SBITP#1{\hbox to\hsize{\tenpoint \baselineskip=12pt
        \hfil\vtop{
        \hbox{\strut SB-ITP-96-#1}
}}}
\def\TUPT#1{\hbox to\hsize{\tenpoint \baselineskip=12pt
        \hfil\vtop{
        \hbox{\strut TUPT-96-#1}
}}}
\def\PRINCETON{
\centerline{${}^1$Physics Department, Princeton University}
\centerline{Princeton, New Jersey 08544}}
\def\STONY{
\centerline{${}^2$Institute for Theoretical Physics, %
State University of New York}
\centerline{Stony Brook, New York 11794}}
\def\TUFTS{
\centerline{${}^3$Institute for Cosmology, Physics Department, %
Tufts University}
\centerline{Medford, Massachusetts 02155
}}

\def\TITLE#1{\vskip .0in \centerline{\fourteenpoint #1}}
\def\AUTHOR#1{\vskip .1in \centerline{#1}}

\def\ABSTRACT#1{\vskip .1in \vfil \centerline{\twelvepoint
\bf Abstract}
   #1 \vfil}
\def\ENDTITLEPAGE{\vfil\eject\pageno=1}
\def\overleftrightarrow#1{\vbox{\ialign{##\crcr
$\leftrightarrow$\crcr\noalign{\kern-1pt\nointerlineskip}
$\hfil\displaystyle{#1}\hfil$\crcr}}}

\def\part{\partial}

\hfuzz=5pt
\tolerance=10000
\TITLEPAGE
\PUPT{1666}
\SBITP{68}
\TUPT{06}
\vskip 1in
\TITLE{Primordial Gravitational Waves From Open Inflation}

\hskip 1in
{\baselineskip=14pt

\AUTHOR{Martin Bucher\footnote\dagger{Current address: Institute
for Theoretical Physics, State University of New York,
Stony Brook, New York 11794. 
E-mail: bucher@insti.physics.sunysb.edu}${}^{1,2}$
and J.D. Cohn\footnote\ddagger{E-mail: 
jdc@cosmos2.phy.tufts.edu}${}^{3}$}
\vskip 14pt
\PRINCETON
\vskip 14pt
\STONY
\vskip 14pt

\TUFTS

}

\nobreak
\ABSTRACT{We calculate the spectrum of
gravitational waves generated
during inflation in open $(\Omega _0<1)$ inflationary models. In
such models an initial epoch of old inflation solves the
horizon and flatness problems, and during this first epoch of
inflation the quantum state of the graviton field rapidly approaches
the Bunch-Davies vacuum. Then old inflation ends by the nucleation
of a single bubble, inside of which there is a shortened epoch
of slow-roll inflation giving $\Omega _0<1$ today. In this paper
we re-express the Bunch-Davies vacuum for the graviton field
in terms of the hyperbolic modes inside the bubble and propagate
these modes forward in time into the present era. We derive the
expression for
the contribution from these gravity waves
to the cosmic microwave background anisotropy including the effect of
a finite energy difference across the bubble wall. 
[PACS numbers 98.80Bp, 98.80Cq]}

\rightline{[January 1997]}

\ENDTITLEPAGE
\eject
\def\xz{\xi ; \zeta}

\REF\gwa{L. Grishchuk, ``The Amplification of Gravitational
Waves and Creation of Gravitons in the Isotropic Universe,"
Lett. Nuovo Cimento {\bf 12,} 60 (1975), Erratum: Ibid. {\bf 12,}
432 (1975); L. Grishchuk, ``Amplification of Gravitational Waves
in an Isotropic Universe," Zh. Eksp. Teor. Fiz. {\bf 67,} 825 (1974).
[Translation: Sov. Phys. JETP {\bf 40,} 409 (1975)].}

\REF\gwb{A. Starobinsky, ``A New Type of Isotropic Cosmological
Models Without Singularity," Phys. Lett. {\bf B91,} 99 (1980);
A. Starobinsky, ``Relict Gravitation Radiation Spectrum and
Initial State of the Universe," Pisma Zh. Eksp. Teor. Fiz.
{\bf 30,} 719 (1979) [Translation: JETP Lett. {\bf 30,}
682 (1979)].}

\REF\gwc{L. Abbott and M. Wise, ``Constraints on Generalized
Inflationary Cosmologies," Nucl Phys. {\bf B244,} 541 (1984);
L. Abbott and M. Wise, ``Anisotropy of the Microwave Background
in the Inflationary Cosmology," Phys. Lett. {\bf B135,} 279 (1984).}

\REF\abbott{L. Abbott and R. Schaefer, ``A General,
Gauge-Invariant Analysis of the Cosmic Microwave Anisotropy,"
Ap. J. {\bf 308,} 462 (1986).}

\REF\ah{ L. Abbott and D. Harari,
``Graviton Production in Inflationary Cosmology,''
Nucl. Phys. {\bf B264,} 487 (1986).}

\REF\gwd{M. White, ``Contribution of Long Wavelength
Gravitational Waves to the Cosmic Microwave Background
Anisotropy," Phys. Rev. {\bf D46,} 4198 (1992);
B. Allen and S. Koranda, ``CBR Anisotropy from Primordial
Gravitational Waves in Inflationary Cosmologies," Phys. Rev. {\bf D50,}
3713 (1994).}

\REF\gwz{L. Krauss and M. White, ``Grand Unification, Gravitational
Waves, and the Cosmic Microwave Background Anisotropy,"
Phys. Rev. Lett. {\bf 69,} 869 (1992);
R. Crittenden, J.R. Bond, R. Davis, G. Efstathiou, and P.
Steinhardt, ``The Imprint of Gravitational Waves on the Cosmic
Microwave Background," Phys. Rev. Lett. {\bf 71,} 324 (1993).}

\REF\gott{J.R. Gott, III, ``Creation of Open Universes from de Sitter Space,"
Nature {\bf 295,} 304 (1982); J.R. Gott and T. Statler, ``Constraints on the
Formation of Bubble Universes," Phys. Lett. {\bf 136B,} 157 (1984);
J.R. Gott, ``Conditions for the Formation of Bubble Universes," in
E.W. Kolb et al., Eds., {\it Inner Space/Outer Space,} (Chicago:
University of Chicago Press, 1986).}

\REF\cd{S. Coleman and F. de Luccia, ``Gravitational Effects on and of
Vacuum Decay," Phys. Rev. {\bf D21,} 3305 (1980).}

\REF\bgt{M. Bucher, A.S. Goldhaber, and N. Turok, ``An Open Universe
{}From Inflation," Phys. Rev. {\bf D52,} 3314 (1995) (hep-ph 94-11206);
M. Bucher and N. Turok, ``Open Inflation with Arbitrary False
Vacuum Mass," Phys. Rev. {\bf D52,} 3314 (1995) (hep-ph 95-03393).}

\REF\sasaki{ M. Sasaki, T. Tanaka, K. Yamamoto, and J. Yokoyama,
``Quantum State During and After Nucleation of an
$O(4)$ Symmetric Bubble,"  Prog. Theor. Phys. {\bf 90,}
1019 (1993); M. Sasaki, T. Tanaka, K. Yamamoto, and J. Yokoyama,
``Quantum State Inside a Vacuum Bubble and Creation of
an Open Universe," Phys. Lett. {\bf B317,} 510 (1993).}

\REF\models{ A. Linde, ``Inflation with
Variable $\Omega$,'' Phys. Lett. {\bf B351,} 99 (1995);
A. Linde and A. Mezhlumian, ``Inflation with $\Omega \ne 1$,''
Phys. Rev. {\bf D52,} 6789 (1995);
L. Amendola, C. Baccigalupi, F. Occhionero,
``Reconciling Inflation with Openness," astro-ph/9504097 (1995);
A. Green and A. Liddle, ``Open Inflationary Universes
in the Induced Gravity Theory,'' astro-ph/9607166 (1996);
J. Garcia-Bellido and A. Liddle, ``Complete Power Spectrum
for an Induced Gravity Open Inflation Model,''
astro-ph/9610183 (1996).}

\REF\ba{B. Allen, ``Vacuum States in de Sitter Space," Phys.
Rev. {\bf D32,} 3136 (1985).}

\REF\norm{M. Sasaki, T. Tanaka, and K. Yamamoto,
``Euclidean Vacuum Mode Functions for
a Scalar Field on Open de Sitter Space,'' Phys. Rev. {\bf D51,} 2979 (1995).
}

\REF\lyth{D. Lyth and E. Stewart, ``Inflationary Density Perturbations
with $\Omega <1,$'' Phys. Lett. {\bf B252,} 336 (1990).}

\REF\ratra{B. Ratra and P.J.E. Peebles, ``Inflation in an Open Universe,"
Phys. Rev. {\bf D52,} 1837 (1995); B. Ratra and P.J.E. Peebles,
``CDM Cosmogony in an Open Universe," Ap. J. Lett. {\bf 432}, L5 (1994).}

\REF\scalar{K. Yamamoto, T. Tanaka, and M. Sasaki,
``Particle Spectrum Created Through Bubble
Nucleation and Quantum Field Theory in the Milne Universe,"
Phys. Rev. {\bf D51,} 2968 (1995).}

\REF\YTSnew{K. Yamamoto, T. Tanaka, and M. Sasaki,
``Quantum Fluctuations and CMB Anisotropies in
One-Bubble Open Inflation Models," Phys. Rev. {\bf D54,} 5031 (1996).}

\REF\scaletwo{ T. Hamazaki, M. Sasaki, T. Tanaka, and
K. Yamamoto, ``Self Excitation of the Tunneling Scalar Field in False
Vacuum Decay,'' Phys. Rev. {\bf D53,}  2045 (1996).}

\REF\wallref{ J. Garriga and A. Vilenkin, ``Perturbations on
Domain Walls and Strings: A Covariant Theory,''
Phys. Rev. {\bf D44,} 1007 (1991); ``Quantum Fluctuations on
Domain Walls, Strings and Vacuum Bubbles,'' Phys. Rev. {\bf D45,}
3469 (1992);
second reference of ref. \models ~and ref. \scaletwo ;
J. Garriga,
``Bubble fluctuations in $\Omega<1$ inflation,'' Phys. Rev. {\bf D54,}
4764 (1996);
J. Garcia-Bellido, ``Density Perturbations from
Quantum Tunneling in Open Inflation,"
Phys. Rev. {\bf D54,}  2473 (1996).}

\REF\jdc{J.D. Cohn, ``Open Universes from Finite Radius Bubbles,"
Phys. Rev. {\bf D54,} 7215 (1996).}

\REF\allen{B. Allen and R. Caldwell, ``Cosmic Background Radiation
Temperature Fluctuations in a Spatially Open Inflationary Universe,"
(unpublished) (1994).}

\REF\robc{R. Caldwell, private communication.}

\REF\mfb{H. Kodama and M. Sasaki, ``Cosmological Perturbation Theory,"
Prog. Theor. Phys. Suppl. {\bf 78,} 1 (1984);
V. Mukhanov, H. Feldman and R. Brandenberger, ``Theory of Cosmological
Perturbations," Phys. Rep. {\bf 215,} 203 (1992).}

\REF\milne{T. Tanaka and M. Sasaki, ``Quantized Gravitational Waves in
the Milne Universe,'' gr-qc/9610060 (1996), and also R. Caldwell,
private communication.}

\REF\tomita{K. Tomita, ``Tensor Spherical and Psuedo-Spherical
Harmonics in Four-Dimensional Spaces," Prog. Theor. Phys. {\bf 68,}
310 (1982).}

\REF\dlaw{D. Lyth and A. Woszczyna, ``Large Scale Perturbations in the Open
Universe," Phys. Rev. {\bf D52,} 3338 (1995).}

\REF\hawking{S.W. Hawking and G.F.R. Ellis, {\it The Large-Scale
Structure of Space-Time,} (Cambridge: Cambridge University Press,
1973).}

\REF\yssimple{M. Sasaki and T. Tanaka,
``Can the Simplest Two-Field Model of Open Inflation
Survive''? astro-ph/9605104 (1996).}

\REF\sasakinew{T. Tanaka and M. Sasaki,
``Quantum State During and After O(4) Symmetric Bubble
Nucleation with Gravitational Effects," Phys. Rev. {\bf D50,}
6444 (1994).}

\REF\parke{S. Parke, ``Gravity and the Decay of the False Vacuum,"
Phys. Lett. {\bf 121B,} 313 (1983).}

\REF\sachs{R. Sachs and A. Wolfe, ``Perturbations of a Cosmological
Model and Angular Variation of the Microwave Background,"
Ap. J. {\bf 147,} 73 (1967).}

\REF\cmbrev{M. White, D. Scott, and J. Silk, ``Anisotropies in the
Cosmic Microwave Background," Ann. Rev. Astron. and
Astrophys. {\bf 32,} 319 (1994).}

\REF\baclosed{B. Allen, R. Caldwell, and S. Koranda, ``CBR
Temperature Fluctuations Induced by Gravitational Waves
in a Spatially Closed Inflationary Universe,"
Phys. Rev. {\bf D51,} 1553 (1995).}

\REF\deg{M. R. de Garcia Maia and J.A.S. Lima,
``Graviton Production in Elliptical and Hyperbolic
Universes," Brown-HET-1047, gr-qc/9606032 (1996).}

\REF\wh{ W. Hu and M. White, to appear.}
\REF\allentwo{B. Allen and R. Caldwell, forthcoming paper.}

\REF\sasakifive{M. Sasaki et al., astro-ph/9701053.}

\REF\bander{M. Bander and C. Itzykson, ``Group Theory and
the Hydrogen Atom (II)," Rev. Mod. Phys. {\bf 38,} 346 (1966).}

\REF\birrell{ N. Birrell and P. Davies, {\it Quantum Fields in
Curved Space,} (Cambridge: Cambridge University Press, 1982), and
references therein.}

\REF\thorne{K. Thorne, ``Multipole Expansions of Gravitational
Radiation," Rev. Mod. Phys. {\bf 52,} 299 (1980).}

\REF\biten{B. Allen, ``Maximally Symmetric Spin Two Bitensors
on $S^3$ and $H^3$," Phys. Rev. {\bf D51,} 5491 (1995).}

\REF\gerlach{U. H. Gerlach, ``Thermal Ambience of Expanding Event
Horizon in Minkowski Spacetime," Phys. Rev. {\bf D28,} 761 (1983).}

\REF\redmount{I. Redmount and S. Takagi, ``Hyperspherical
Rindler Space, Dimensional Reduction, and de Sitter Space
Scalar Field Theory," Phys. Rev. {\bf D37,} 1443 (1988).}

\def\K{{\cal K}}
\def\gtorder{\mathrel{\raise.3ex\hbox{$>$}\mkern-14mu
             \lower0.6ex\hbox{$\sim$}}}
\def\ltorder{\mathrel{\raise.3ex\hbox{$<$}\mkern-14mu
             \lower0.6ex\hbox{$\sim$}}}
\def\Re{{\cal R}{\it e}}

\chapter{Introduction}

It has been well known for quite some time that gravitational
waves are generated during inflation and give an observable,
and sometimes substantial, contribution to the cosmic microwave
background anisotropy [\gwa -\gwz ].
For flat $(\Omega _0=1)$ inflation the
spectrum of gravitational waves generated and their observational
consequences have been studied quite extensively. However, for
open $(\Omega _0<1)$ inflationary models in which our entire
observable universe lies inside a single bubble
[\gott,\cd,\bgt,\sasaki], there has been
no complete calculation of the gravitational waves generated. In this
paper we present such a calculation.

Gravitational waves from inflation result from the stretching of
quantum vacuum fluctuations of the linearized graviton field
to superhorizon scales. Following a given mode of
fixed co-moving wavenumber $k,$ one finds that at
early times its physical wavelength $\lambda =
a(t)\cdot (2\pi )/k$ is much smaller than the
Hubble length $\ell _H=H^{-1}$ (i.e., the mode is well within
the {\it horizon}). This implies that for determining a physically
reasonable {\it vacuum state} for the mode at early times, one may
ignore the expansion of the universe and match onto
the usual flat Minkowski space vacuum for the graviton field. Once the
correct quantum vacuum state has been determined at early times,
to continue these modes to later times become a mathematically well-defined
exercise in classical field theory, which involves
propagating the {\it positive} frequency modes---those
associated with annihilation operators of the
vacuum---forward in time, through the end of inflation
into the present epoch. As a mode crosses the horizon
its amplitude becomes frozen in. The process of
generating gravitational waves during inflation
is quite analogous to the process of generating
scalar fluctuations during inflation. There is, however,
an important difference. The amplitude
of the gravitational waves does not depend on the
slope of the potential; rather only the overall height of
the potential, or equivalently the expansion rate $H$ during
inflation, is relevant. As a first approximation for calculating
the gravitational waves, $H$ may be
regarded as fixed during the relevant epoch of inflation.

For open inflation identifying the correct initial
conditions for the linearized graviton field is not
as straightforward as for the flat case, because
the underlying spacetime geometry is more complicated.
The simplest case involves one matter field, the inflaton
$\phi$,
and minimal coupling to gravity.
More complicated models such as found in [\models]
have similar properties for the gravity wave
calculations done here, so
the simple one field model is used in the following
description.  (For a more detailed
discussion of single bubble inflation see [\bgt].)

In the one field models, there is an initial epoch of old inflation
during which $\phi $ is stuck in a false vacuum
with $\phi =\phi _{fv}.$
During this epoch,
the spacetime geometry approaches that of pure de Sitter space,
characterized by an expansion rate $H_{fv}$ where
$H_{fv}^2=(8\pi G/3)V[\phi _{fv}].$ During this initial
epoch of old inflation the graviton field is driven
to the vacuum state.  This determines the initial conditions
using the same considerations as for the flat case described
above. Then old inflation ends through the nucleation of a
bubble, which expands roughly at the speed of light.
The preferred time slicing inside the
light cone of the bubble center corresponds to a
spatially open universe.
Inside the bubble the inflaton field first slowly rolls
down a rather flat part of the potential, giving a
shortened epoch of slow-roll inflation inside the bubble.
Later inside the bubble, the inflaton field rolls
more quickly and the usual reheating occurs, converting the
vacuum energy of the inflaton field into radiation and
matter.

The coordinate chart with the line element
$$ds^2=-dt^2+a^2(t)\cdot \bigl[ d\xi ^2+\sinh ^2[\xi ]~
d\Omega _{(2)}^2\bigr] ,\eqn\aaa$$
describes an expanding Friedman Robertson Walker
universe with spatially uniform
negative spatial curvature.
Hyperbolic open coordinates
(rather than the {\it flat} coordinates, to be described later)
are the natural coordinate
choice in the presence of the bubble wall, which is
why the interior of a bubble is an open universe [\gott, \cd].
For de Sitter space and these
hyperbolic coordinates, $a(t) = \sinh [t]$, and
$$ds^2= -dt_h^2+\sinh ^2[t_h]\cdot \Bigl[ d\xi ^2+\sinh ^2[\xi ]~
d\Omega _{(2)}^2\Bigr] \eqn\pbc $$
where $(0\le t_h<+\infty ).$
The hyperbolic coordinate chart
has an unphysical coordinate singularity
at $t=0,$ and to determine initial conditions it
is necessary to consider a larger region of spacetime than
that covered by the open coordinates with the line element
\aaa .

The bubble nucleation process underlying open inflation
is sketched in Fig.~1, with the dashed
lines indicating the surfaces on which inflaton field is
constant. Roughly speaking, the forward light cone of the
materialization center $M,$ which we shall call region I
and which is covered by the coordinate chart just described,
may be considered the bubble interior. Regions II and III
cover the spacetime prior to bubble nucleation and the
part of spacetime into which the bubble expands, at a speed
approaching the speed of light.

\vskip -.5in
\def\Onne{1}
\LoadFigure\Onne{\baselineskip 13 pt
\noindent\narrower\ninerm
{\ninebf ---Spacetime Diagram for Open Inflation.}
Fig.~1(a) shows a spacetime diagram for bubble nucleation.
The double-dashed vertical line to the left
indicates an $r=0$ coordinate singularity. Time flows
upward and the horizontal axis represents a radial
coordinate. On the surfaces represented by dashed curves
the inflaton field is constant. The lower
portion of the diagram (with $t<0$) represents the nucleation
of a critical bubble, a classically forbidden
Euclidean process. For $t>0$ the bubble expands
classically, at a speed approaching that of
light. The classical expanding bubble evolution is $SO(3,1)$ symmetric.
In Fig.~1(b) the hyperbolic coordinates that
maximally exploit the $SO(3,1)$ symmetry of the
expanding bubble solution are sketched. Spacetime
is divided into three hyperbolic coordinate patches.
The light cones separating these regions represent
unphysical coordinate singularities of a character
similar to that of the Schwarzschild horizon.}
{\epsfysize 5truein}{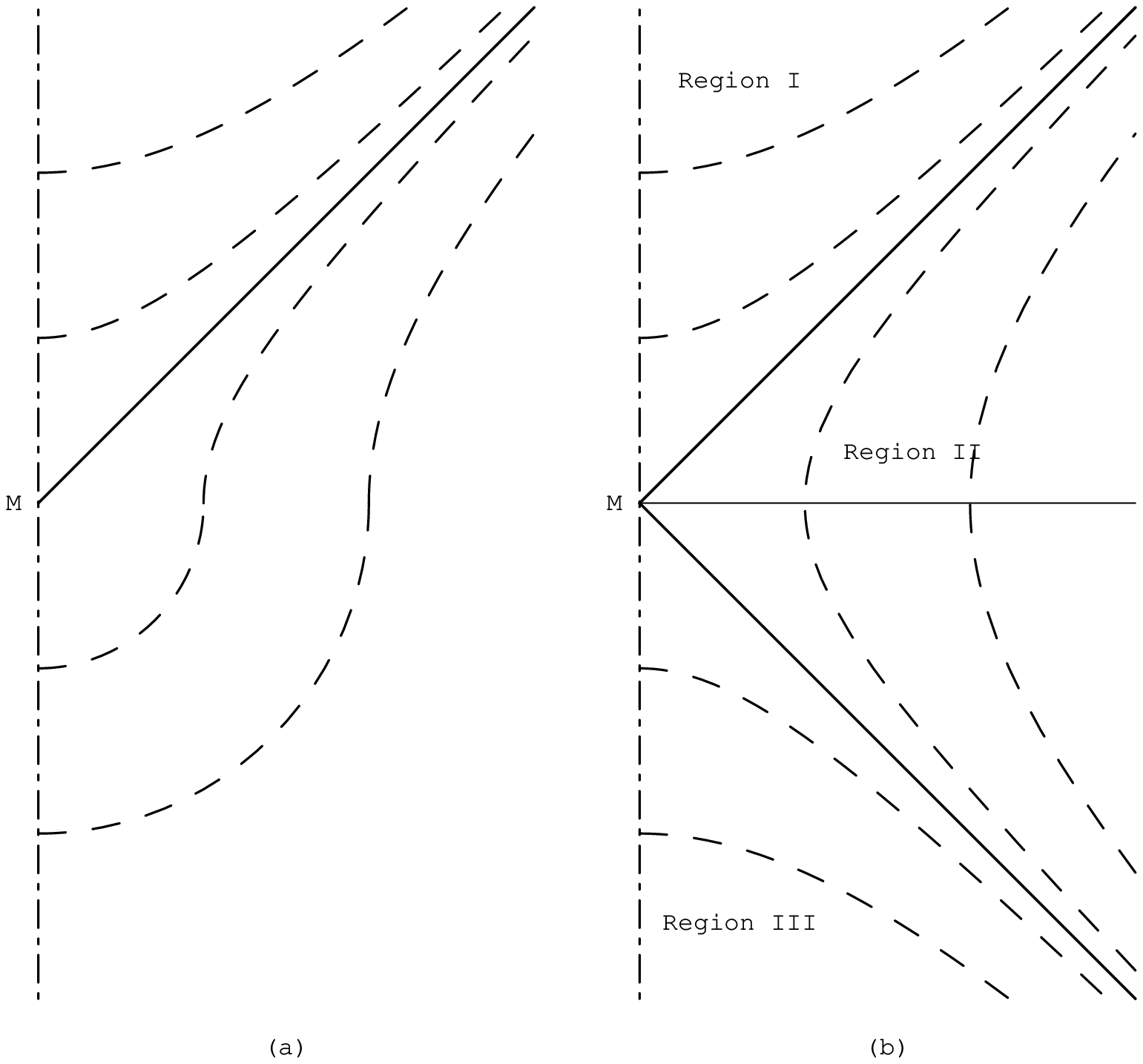}{}

For future reference, in Fig.~2, we present the conformal
diagram for all of maximally extended de Sitter space. In
addition to the regions already mentioned, there
also exist regions IV and V, which are the past and future
light cones of $\bar M,$ the antipodal point of the apex of region I.
For a discussion of the global structure of the de Sitter vacuum
see [\ba,\norm].

\vskip -.5in
\def\Two{2}
\LoadFigure\Two{\baselineskip 13 pt
\noindent\narrower\ninerm
{\ninebf ---Hyperbolic Coordinates for Maximally Extended
de Sitter Space.} A conformal diagram for
all of maximally extended de Sitter space is shown here.
$\bar M$ is the antipodal point of $M.$ The hyperbolic coordinates
that exploit the symmetry of the $SO(3,1)$ subgroup of the
full de Sitter group $SO(4,1)$ that leaves invariant $M$ (and $\bar M$
as well) divides spacetime into the five indicated
coordinate patches.}
{\epsfysize 4.5truein}{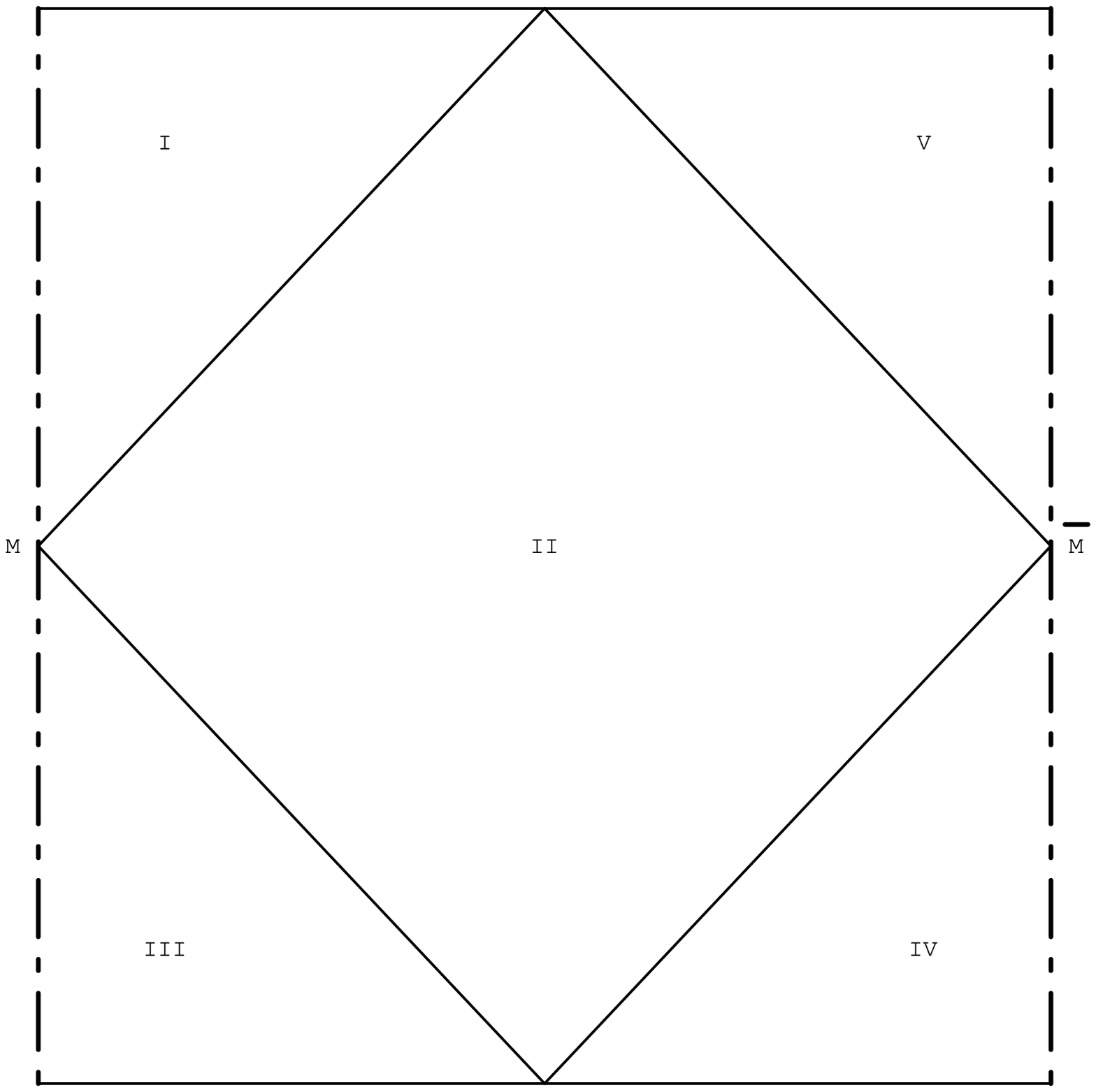}{}

The subject of {\it scalar} perturbations in open inflation
has been studied extensively in recent years. Lyth and
Stewart [\lyth ] and Ratra and Peebles [\ratra ] calculated the scalar
perturbations in open inflation assuming conformal boundary
conditions for the vacuum of the inflaton field as $t\to 0$
in region I. Bucher, Goldhaber, and Turok [\bgt ] presented the
first computation of the scalar perturbations in open inflation
using the Bunch-Davies vacuum for the inflaton field outside the
bubble during the prior epoch of
old inflation as initial conditions and
propagating the scalar modes through the bubble wall into the open
universe. Yamamoto, Sasaki, and Tanaka obtained essentially the
same results using Euclidean methods [\scalar , \YTSnew ].
Many recent calculations of scalar field
perturbations, for example
[\models ,\scaletwo , \wallref , \YTSnew, \jdc ],
take
into account additional effects such as
finite bubble size, varying bubble
wall profile, fluctuations in the bubble wall, {\it et cetera}.

The subject of gravitational waves generated during open
inflation was previously investigated by B. Allen and
R. Caldwell in an unpublished manuscript [\allen ].
Within their approximations they found an infrared
divergence in the multipole moments of the cosmic microwave
background anisotropy (CMB).
In their computation,
in order to simplify the calculation, at early times
a flat spacetime geometry (i.e., that of Minkowski space)
is assumed, and to improve the infrared behavior of the
graviton field during this early epoch, the graviton field is
given a small mass $m_g,$ which at the end of the calculation
taken to approach zero. Later on, inside region
I on a hyperboloid of constant time with respect to the region I
coordinates, the spacetime geometry is taken to change
discontinuously to that of de Sitter space. Subsequently,
inflation inside the bubble ends, so that the scale factor
$a(\eta )$ becomes governed by a radiation-matter equation of
state, and the evolution of the graviton field is computed to
take into account this change.
This is a situation different from open inflation and
violates the Einstein equations for the background solution.

In section IV of this paper we derive the
Bunch-Davies vacuum for the graviton field in de Sitter
space in terms of the hyperbolic region I coordinates,
assuming de Sitter space everywhere and a vanishing graviton mass.
The Bunch-Davies vacuum
for the graviton field in de Sitter space, expressed in 
terms ofthe
spatially flat coordinate slicing, is used to
express the Bunch-Davies vacuum
for the graviton field in de Sitter space in terms of
the hyperbolic coordinates.  The latter are the natural
coordinates for studying perturbations in an open
universe. While in principle it should be possible
to compute directly the transformation between the flat
and the hyperbolic modes, in practice this
transformation has proven algebraically intractable.
Instead, we characterize linear combinations of modes of
{\it positive} frequency with respect to the flat coordinates
(which is the same as {\it positive} frequency with respect to
the Bunch-Davies vacuum) in terms of analytic properties in the
complex plane. In this way linear combinations of hyperbolic modes
of purely {\it positive} frequency  with respect to
the Bunch-Davies vacuum may be constructed without explicitly
expressing these combinations in terms of the flat modes.
The basic approach is somewhat analogous to the Euclidean methods
used by Sasaki et al.~for scalar modes [\norm ], except that
here, rather than using a Euclidean principle
as the starting point, we determine the
vacuum in terms of the flat modes, using the
complex plane as a mathematical tool.

Our result for the Bunch-Davies vacuum for the graviton field essentially
coincides with that found by Allen and Caldwell. There
are some minor differences: whereas we obtain a mixed
state directly, Allen and Caldwell obtain a pure state,
which in the limit $m_g\to 0$ mocks our mixed
state. This does not alter the divergence of
the CMB multipole moments.

One caveat in computing the CMB moments relates to gauge fixing.
In the Sachs-Wolfe formula for the CMB anisotropy the gravity waves
contribute only in the integral along the line of sight---there is
no contribution from the {\it tensor} modes from the last scattering
surface. Since linearized gauge transformations are {\it scalar}
and {\it vector}, one would think that the {\it tensor} part
is gauge invariant.%
\footnote\ddagger{We define {\it scalar,} {\it vector,}
and {\it tensor} here such that a vector that can be
expressed as the spatial gradient of a scalar is
regarded as a {\it scalar,} a tensor that may be expressed
as spatial derivatives acting on a {\it vector} is
regarded as a {\it vector,} etc.}
However, gauge transformations can move
the last scattering surface, and thus change the {\it intrinsic}
contribution in the Sachs-Wolfe formula. If gravity wave
modes mix with pure gauge modes, cancellations may occur [\robc ].
Linearized gravity wave modes typically are taken to
satisfy the synchronous gauge
condition $\hat t^\mu h_{\mu \nu }=0$ where $\hat t^\mu $
points along the time direction of the preferred coordinate
system. In de Sitter space
these conditions do not coincide for the flat and for the region I
hyperbolic coordinatizations. These issues have been
investigated for hyperbolic coordinates in flat Minkowski
space.\refmark{\milne }

In this paper we drop the approximation of no energy
difference across the wall, which is never exactly
the case in the presence of a bubble.  This
requires including finite
critical bubble size as well.
Taking these effects into account removes the
infrared divergence of the CMB multipole moments.
The calculation resembles the
calculations for scalar perturbations in refs.~[\YTSnew, \jdc ].

The organization of this paper as follows. Sections II and
III give the multipole expansion for the pure tensor perturbations
in hyperbolic space. Section II gives the evolution
equation for the graviton field, which is solved by separation
of variables.  In section III we study the properties of
tensor harmonics in three-dimensional hyperbolic space,
first written down explicitly by Tomita [\tomita ].
As mentioned above,
in section IV we identify the Bunch-Davies vacuum for the graviton
field in region I in terms of the hyperbolic modes.
In section V we compute the effect of nonvanishing bubble
size and of nonvanishing energy density difference across the bubble wall.
This result is used in section VI to give the tensor mode
contribution to the CMB anisotropy for an open universe, and finally
section VII concludes.  There are two
appendices containing technical details.  We set $\hbar = 8 \pi G = 1$
throughout, and $H=1$ until section V.

\chapter{Gravitational Waves in an Open Universe}

Gravitational waves are fluctuations about a background
metric. The metric can be written as a background metric
$g^B_{\mu \nu }$ plus a small perturbation:
$$g_{\mu \nu }=g^B_{\mu \nu }+\hat h_{\mu \nu }\eqn\xxaa$$
where we set
$\hat h_{00}=0,$ $\hat h_{0i}=0,$ $\hat h^{~i}_i =0,$ and
${\hat h_{ij}}^{~~\vert j}=0.$ These conditions
require the {\it scalar} and {\it vector} perturbations to vanish
and fix the gauge as well.
The unperturbed spatial metric $\gamma _{ij}$ (with line
element $d\bar s^2=d\xi ^2+\sinh ^2[\xi ]d\Omega _{(2)}^2$) is used to
raise and lower roman indices, and the vertical line indicates
the covariant derivative induced by $\gamma _{ij}.$

For the background corresponding to the natural coordinates
in the interior of the bubble center's light cone, the
background metric is given in eqn. \pbc.  Hyperbolic
conformal time
$$ \eta =\ln \Bigl[ \tanh [t_h/2]\Bigr] \; ,
\eqn\pbd $$
will be used primarily in the following, in which the
line element is
$$ ds^2={ a^2[\eta]}\cdot
\Bigl[-d \eta^2 + d\xi ^2+\sinh ^2[\xi ]~d\Omega _{(2)}^2\Bigr] ,
\eqn\pbe $$
where $-\infty < \eta < 0$.

The condition that the first order perturbation of the Ricci tensor
vanishes $\delta R^{(1)}_{\mu \nu }=0$
gives the equation of motion\refmark{\mfb }
$$g^{\alpha \beta }\nabla _\alpha \nabla _\beta \hat h_{\mu \nu }
+2R^{(B)}_{\alpha \mu \beta \nu }\hat h^{\alpha \beta }=0.
\eqn\xxa$$
This may be rewritten as
$$\left[ D_\eta ^2-\nabla _{(3)}^2+2\K \right]
h_{ij}(\xi ,\theta ,\phi ,\eta )=0,
\eqn\lya$$
where $\K $ is the spatial curvature, with $\K =-1$ for a
hyperbolic universe, and we have used
$R^{(B)}_{ijkl}=\K (\gamma _{ik}\gamma _{jl}-\gamma _{jk}\gamma
_{il}).$
Using the ansatz
$${\cal T}_i^{~j}({\bf x}, \eta; \zeta ,P,j,m)=n(\zeta )~
\bigl[ {\bf T}^{P,jm}(\xi,\theta,\phi;\zeta)
\bigr] _i^{~j}~T_h(\eta;\zeta),\eqn\lyaa$$
we obtain
$$\eqalign{ &\left[ \nabla _{(3)}^2+(\zeta ^2+3)\right]
{\bf T}^{P,jm}_{ij}(\xi,\theta,\phi;\zeta)
= 0,\cr
&\left[ \partial _\eta ^2+{2a'\over a}%
\partial _\eta +(\zeta ^2+1)\right] T_h(\eta;\zeta)=0.\cr
}\eqn\lyb$$
The normalization
$n(\zeta )$ will be fixed later.

Using properties of the hyperbolic tensor harmonics
that solve eqn.~\lyb a, the gravitational waves
in region I can be expanded as
\footnote\dagger{
It should be pointed out that in addition to the continuous modes
with $0\le \zeta <\infty $ there might also exist some discrete
{\it supercurvature} modes, as have been found for the minimally coupled
scalar field in de Sitter space of mass $m$ when $(m^2/H^2)<2.$
See refs.~[\norm ] and [\dlaw ] for a discussion.  The one supercurvature
mode
which would be expected to appear here by direct analogy has zero
contribution to the CMB because of its lack of time dependence.}
$$\eqalign{
\hat{h}_{rs}(\eta,\xi,\theta,\phi) &= \sum_{jmP} \int _0^\infty
d \zeta ~n(\zeta)~{\bf T}^{P,jm}_{ij}(\xi,\theta,\phi;\zeta)~
T_h(\eta;\zeta) ~\hat{a}_{P,j,m}(\zeta ) + \; h.c.  \cr
&=\sum _{jmP} \int _0^\infty
d \zeta ~~{\cal  T}^{P,jm}_{ij}(\xi,\theta,\phi,\eta ;\zeta)~
\hat{a}_{P,j,m}(\zeta ) + \; h.c.  \cr }
\eqn\decomp$$
Here the operators $\hat a_I(\zeta ,P,j,m),$
$a_I^\dagger (\zeta ,P,j,m)$ satisfy the canonical commutation relations
$[\hat a_I(\zeta ,P,j,m), ~\hat a_I(\zeta ',P',j',m')]=
[\hat a_I^\dagger (\zeta ,P,j,m), ~\hat a_I^\dagger (\zeta ',P',j',m')]
=0$ and $[\hat a_I(\zeta ,P,j,m), ~\hat a_I^\dagger (\zeta ',P',j',m')]=
\delta (\zeta -\zeta )~\delta _{PP'}~\delta _{jj'}~\delta _{mm'}$
where $\zeta ,\zeta '>0.$

 The spatial
tensor harmonics and $n(\zeta)$ are discussed in the following
section.
Here the time dependence for the hyperbolic tensor modes
and the flat tensor modes is found.
For open de Sitter space,
$a(\eta )=-1/\sinh [\eta ]$.  Thus for
hyperbolic tensor modes eqn.~\lyb b becomes
$$T^{\prime \prime }_h-2\coth [\eta ]T'_h+(\zeta ^2+1)T_h=0,
\eqn\jjc $$
replacing
$-\nabla ^2_{(3)}$ and $\K $ with $(\zeta ^2+3)$
and $-1,$ respectively. Transforming to the dependent variable
$T=\sinh ^2[\eta ]\cdot F,$ one may recast eqn.~\jjc ~into the form
$$F^{\prime \prime }+2\coth [\eta ]F'+(\zeta ^2+1)F-
{2\over \sinh ^2[\xi ]}F=0, \eqn\jjd $$
which is identical to the equation for the spatial hyperbolic
radial functions with orbital angular momentum
$l=1.$ [See for example ref.~[\bgt ], eqn.~(5.21).]
It follows that
$$T_h(\eta ;\zeta )=\bigl\{ i\zeta \sinh [\eta ]+\cosh [\eta ]\bigr\} \cdot
e^{-i\zeta \eta }\eqn\jje$$
where $\zeta $ is allowed to take both signs.

For the flat tensor modes, which will be needed in order to find
the Bunch-Davies vacuum, the
time evolution equation becomes
$$T^{\prime \prime }_f-{2\over \bar \eta }T'_f+\omega ^2T_f=0,
\eqn\jjh$$
where we replace $-\nabla ^2 $ and $\K $ with $\omega ^2 $
and $0,$ respectively. With the substitution $T=\eta_f ^{3/2}
\cdot H,$ eqn.~\jjh ~ becomes the Bessel equation of order
$\nu =3/2$ whose solutions $H_{3/2}^{(+)}(\omega \eta_f )$
and $H_{3/2}^{(-)}(\omega \bar \eta )$ are proportional to
the spherical Bessel functions $h_1^{(+)}(\omega \eta_f )$
and $h_1^{(-)}(\omega \eta_f )$
multiplied by $\eta_f ^{1/2}.$ Consequently,
$$T_f(\eta_f ;\omega )=
{1\over \omega ^2}\cdot
[1+i\omega \eta_f ]\ e^{-i\omega \eta_f }\eqn\jji$$
where $\omega $ is allowed to take both signs.

\chapter{Hyperbolic Tensor Harmonics}

We now turn to computing and normalizing the hyperbolic tensor
harmonics, which satisfy the equation
$$h_{ij~~~\vert k}^{~~\vert k}+(\zeta ^2+3)h_{ij}=0.\eqn\bac$$
The offset in $(\zeta ^2+3)$ is chosen for later
convenience.
The pure tensor character of these modes requires that
they satisfy
the conditions of tracelessness
$$h_i^{~i}=0,\eqn\baa$$
and transversality
$$h_{ij}^{~~\vert j}=0.\eqn\bab$$
Here the roman letters $(i,j=1,2,3)$ indicate spatial indices.

Since eqns.~\bac --\bab ~are invariant under
rotations and spatial inversion about $\xi =0,$
multipole solutions may be classified according to their
angular momentum quantum numbers ${\bf J}^2,$ $J_3,$
and their parity $\pi .$  Parity is either
{\it electric} with $\pi =(-)^j$ or {\it magnetic} with $\pi =(-)^{j+1},$
denoted by $P=E$ and $P=M,$ respectively.
Fixing $j,$ $m,$ and $P,$ we write down the most general symmetric
tensor field with these quantum numbers.  This restricts
the angular dependence to a few terms but does not specify
the radial dependence.  Imposing
eqns.~\bac --\bab ~and solving for the $\xi $-dependence
gives a solution for each $\zeta >0$ and $(j,m,P)$ for $j\ge 2$ unique up to
an overall normalization.  (There are no monopole $(j=0)$ or
dipole $(j=1)$ modes.)
The solution to these conditions has been found by
Tomita[\tomita].

The tensor field with electric parity has the form
$$\eqalign{
{\bf \tilde{T}}^{E, jm}(\xi ,\theta ,\varphi ; \zeta )&=~~
F_j(\xi ; \zeta  )~~({\bf e}^\xi\otimes {\bf e}^\xi )~~
Y_{jm}(\theta ,\varphi )\cr
&~~+ G_j(\xi ; \zeta )~\delta _{\tilde a \tilde b}~
({\bf e}^{\tilde a}\otimes {\bf e}^{\tilde b})~
Y_{jm}(\theta ,\varphi ) \cr
&~~+ H_j(\xi ; \zeta )~(
{\bf e}^{\tilde a}\otimes {\bf e}^{\xi }
+{\bf e}^{\xi }\otimes {\bf e}^{\tilde a})~
\tilde \nabla _{\tilde a}Y_{jm}(\theta ,\varphi ) \cr
&~~+I_j(\xi ; \zeta )~({\bf e}^{\tilde a}\otimes {\bf e}^{\tilde b})~
\tilde \nabla _{\tilde a}\tilde \nabla _{\tilde b}
Y_{jm}(\theta ,\varphi )\cr }
\eqn\bad$$
where $(\tilde a,\tilde b =1,2)$ indicate angular indices and
$\tilde \nabla $ indicates an $S^2$ (rather than an $H^3$) covariant
derivative.  As a result,  $\delta _{\tilde a \tilde b}~
({\bf e}^{\tilde a}\otimes {\bf e}^{\tilde b})
=({\bf e}^\theta \otimes {\bf e}^\theta +\sin ^2\theta ~{\bf e}^\varphi
\otimes {\bf e}^\varphi ).$
The basis functions are
${\bf e}^\xi =d\xi ,$ ${\bf e}^\theta =d\theta ,$
and ${\bf e}^\varphi
=d\varphi .$ Note that this is not a {\it vielbein} (normalized) basis.

We use
$$\nabla ^2=D_\xi ^2+2\coth [\xi ]D_\xi +
{1\over \sinh ^2[\xi ]}\cdot \left[
D_\theta ^2+\cot [\theta ]D_\theta +{1\over \sin ^2\theta }
D_\varphi ^2 \right] \eqn\bada$$
and
$$\eqalign{
D_\xi {\bf e}^\xi &= 0,\cr
D_\xi {\bf e}^\theta &= -\coth [\xi ]{\bf e}^\theta ,\cr
D_\xi {\bf e}^\varphi &= -\coth [\xi ]{\bf e}^\varphi,\cr
D_\theta {\bf e}^\xi &= +\sinh [\xi ]\cosh [\xi ]{\bf e}^\theta ,\cr
D_\theta {\bf e}^\theta &= -\coth [\xi ]{\bf e}^\xi .\cr }\qquad
\eqalign{
D_\theta {\bf e}^\varphi &= -\cot [\theta ]{\bf e}^\varphi ,\cr
D_\varphi {\bf e}^\xi &= +\sinh [\xi ]\cosh [\xi ]\sin ^2[\theta ]
{\bf e}^\phi ,\cr
D_\varphi {\bf e}^\theta &=+\sin [\theta ]\cos \theta ]{\bf e}^\varphi ,\cr
D_\varphi {\bf e}^\varphi &= -\cot [\theta ]{\bf e}^\theta -\coth [\xi ]
{\bf e}^\xi ,\cr &\cr }
\eqn\badb$$

We now impose the constraints of transversality and tracelessness.
For transversality,
taking the divergence of ${\bf T}^{jm, E}$ gives
$$\eqalign{
\nabla \cdot T_{jm} =
&\left[ {\partial F\over \partial \xi }+2\coth [\xi ]F(\xi )
-{2\coth [\xi ]\over \sinh ^2[\xi ]}G(\xi )-
{j(j+1)\over \sinh ^2[\xi]}H(\xi ) \right.  \cr
& \left. +{j(j+1)\coth [\xi ]\over \sinh ^2[\xi ]}I(\xi )
\right] Y_{jm}(\Omega )\times {\bf e}^\xi \cr
&+\left[
{G(\xi )\over \sinh ^2[\xi ]}
+{\partial H\over \partial \xi }+2\coth [\xi ]H(\xi )
-{1\over \sinh ^2[\xi ]}\cdot [j(j+1)-1]I(\xi )
\right] \cr
&~~~~~~~~~\times \left[
{\bf e}^\theta
{\partial Y_{jm}\over \partial \theta }+
{\bf e}^\varphi {\partial Y_{jm}\over \partial \varphi }\right]
=0.}\eqn\badza$$
Both terms must individually vanish. Likewise, taking the
trace and asking it to vanish gives the condition
$$ T^i_i = F(\xi )+{2G(\xi )\over \sinh ^2[\xi ]}-{j(j+1)I(\xi )
\over \sinh ^2[\xi ]}=0.
\eqn\badzb$$

Thus transversality and tracelessness give
$$\eqalign{
H_j(\xz  )&={\sinh ^2[\xi ]\over j(j+1)}\cdot
\left[ {\partial F_j(\xz )\over \partial \xi }+3\coth [\xi ]F_j(\xz  )
\right] ,\cr
I_j(\xz  )&={\sinh ^2[\xi ]\over (j+2)(j-1)}\cdot
\left[ 2\left( {\partial H_j(\xz)\over \partial \xi }+
2\coth [\xi ]~H_j(\xz  )\right) -F_j(\xz  )\right] ,\cr
G_j(\xz  )&={1\over 2}\biggl[ j(j+1)I_j(\xz  )-\sinh ^2[\xi]F_j(\xz  )\biggr]
.\cr } \eqn\badzc$$

The Laplacian in eqn.~\bada ~acting on the various components of
${\bf \tilde{T}}^{E,jm}$ in eqn.~\bad ~gives
$$\eqalign{
&\nabla ^2\left[ F(\xi )Y_{jm}(\Omega )
\cdot ({\bf e}^\xi \otimes {\bf e}^\xi ) \right] \cr
&=
\left[ {\partial ^2F\over \partial \xi ^2}
+2\coth [\xi ]{\partial F\over \partial \xi }
-\left({j(j+1)\over \sinh ^2[\xi ]}+4\coth ^2[\xi ]\right) F(\xi )
\right] Y_{jm}(\Omega)\cdot ({\bf e}^\xi \otimes {\bf e}^\xi ) \cr
&+\left[ 2\cosh ^2[\xi] F(\xi )Y_{jm}(\Omega )\right] \cdot
({\bf e}^\theta \otimes {\bf e}^\theta +\sin ^2\theta {\bf e}^\varphi \otimes
{\bf e}^\varphi ) \cr
&+ \left[ 2\coth [\xi ]F(\xi )\right] \cdot
\left[ ({\bf e}^\xi \otimes {\bf e}^\theta +{\bf e}^\theta \otimes {\bf e}^\xi
)
{\partial Y_{jm}\over \partial \theta }+
({\bf e}^\xi \otimes {\bf e}^\varphi +{\bf e}^\varphi \otimes {\bf e}^\xi )
{\partial Y_{jm}\over \partial \varphi }
\right]
,\cr }
\eqn\badc$$
$$\eqalign{
&\nabla ^2\left[ G(\xi )Y_{jm}(\Omega )
\cdot ({\bf e}^\theta \otimes {\bf e}^\theta +\sin ^2\theta {\bf e}^\varphi
\otimes
{\bf e}^\varphi )\right] \cr
&=\left[ {4\coth ^2[\xi ]\over \sinh ^2[\xi ]}G(\xi )Y_{jm}(\Omega )
\right] \cdot ({\bf e}^\xi \otimes {\bf e}^\xi ) \cr
&+\left[ {\partial ^2G\over \partial \xi ^2}-2\coth [\xi ]
{\partial G\over \partial \xi}-2G-{j(j+1)\over \sinh ^2[\xi ]}G
\right] Y_{jm}(\Omega)\cdot
({\bf e}^\theta \otimes {\bf e}^\theta +\sin ^2\theta {\bf e}^\varphi \otimes
{\bf e}^\varphi )\cr
&-{2\cosh [\xi ]\over \sinh ^3[\xi ]}G(\xi )\cdot
\left[
 ({\bf e}^\xi \otimes {\bf e}^\theta +{\bf e}^\theta \otimes {\bf e}^\xi )
{\partial Y_{jm}\over \partial \theta }+
({\bf e}^\xi \otimes {\bf e}^\varphi +{\bf e}^\varphi \otimes {\bf e}^\xi )
{\partial Y_{jm}\over \partial \varphi }
\right] ,\cr
}\eqn\badd$$
$$\eqalign{
&\nabla ^2\left[ H(\xi )\cdot \left\{
 ({\bf e}^\xi \otimes {\bf e}^\theta +{\bf e}^\theta \otimes {\bf e}^\xi )
{\partial Y_{jm}\over \partial \theta }+
({\bf e}^\xi \otimes {\bf e}^\varphi +{\bf e}^\varphi \otimes {\bf e}^\xi )
{\partial Y_{jm}\over \partial \varphi }
\right\} \right] \cr
&={4j(j+1)\coth [\xi ]\over \sinh ^2[\xi ]}Y_{jm}(\Omega )H(\xi )\cdot
({\bf e}^\xi \otimes {\bf e}^\xi )\cr
&+\left[
{\partial ^2H\over \partial \xi ^2}-2H-4\coth ^2[\xi ]H
-{j(j+1)\over \sinh ^2[\xi ]}H\right] \cr
&~~~\times \left[  (e^\xi \otimes e^\theta +e^\theta \otimes e^\xi )
{\partial Y_{jm}\over \partial \theta }+
({\bf e}^\xi \otimes {\bf e}^\varphi +{\bf e}^\varphi \otimes {\bf e}^\xi )
{\partial Y_{jm}\over \partial \varphi }
\right] \cr
&+4H(\xi )\coth [\xi ]\cdot \biggl[
({\bf e}^\theta \otimes {\bf e}^\theta ){\partial ^2Y_{jm}\over \partial \theta
^2}
+({\bf e}^\varphi \otimes {\bf e}^\varphi )\left(
{\partial ^2Y_{jm}\over \partial \varphi ^2}+\sin \theta \cos \theta
{\partial Y_{jm}\over \partial \theta }\right) \cr
&~~~
+({\bf e}^\theta \otimes {\bf e}^\varphi +{\bf e}^\varphi \otimes {\bf
e}^\theta )
\left( {\partial ^2Y_{jm}\over \partial \theta \partial \varphi }
-\cot \theta {\partial Y_{jm}\over \partial \varphi }\right)
\biggr] ,\cr
}\eqn\bade$$
and
$$\eqalign{
&\nabla ^2\biggl[ I(\xi )\biggl\{
({\bf e}^\theta \otimes {\bf e}^\theta ){\partial ^2Y_{jm}\over \partial \theta
^2}
+({\bf e}^\varphi \otimes {\bf e}^\varphi )\left(
{\partial ^2Y_{jm}\over \partial \varphi ^2}+\sin \theta \cos \theta
{\partial Y_{jm}\over \partial \theta }\right) \cr
&~~~~~~~~~~~~~~~~
+({\bf e}^\theta \otimes {\bf e}^\varphi +{\bf e}^\varphi \otimes {\bf
e}^\theta )
\left( {\partial ^2Y_{jm}\over \partial \theta \partial \varphi }
-\cot \theta {\partial Y_{jm}\over \partial \varphi }\right)
\biggr\} \biggr] \cr
&={-2I(\xi )j(j+1)\coth ^2[\xi ]\over \sinh ^2[\xi ]}
Y_{jm}(\Omega )\cdot ({\bf e}^\xi \otimes {\bf e}^\xi )\cr
&+{2I(\xi ) j(j+1)\over \sinh ^2[\xi ]}Y_{jm}(\Omega)\cdot
({\bf e}^\theta \otimes {\bf e}^\theta +\sin ^2\theta ~{\bf e}^\varphi \otimes
{\bf e}^\varphi )\cr
&+{2[j(j+1)-1]\coth [\xi ]I(\xi )\over \sinh ^2[\xi ]}\cr
&\times ({\bf e}^\xi \otimes {\bf e}^\theta +{\bf e}^\theta \otimes {\bf e}^\xi
)
{\partial Y_{jm}\over \partial \theta }+
({\bf e}^\xi \otimes {\bf e}^\varphi +{\bf e}^\varphi \otimes {\bf e}^\xi )
{\partial Y_{jm}\over \partial \varphi }
\cr
&+\left[ {\partial ^2I\over \partial \xi ^2}-2\coth [\xi ]
{\partial \over \partial \xi }+{6I(\xi )\over \sinh ^2[\xi ]}
-2I(\xi )\coth ^2[\xi ] -{j(j+1)\over \sinh ^2[\xi ]}I(\xi )
\right] \cr
&\times \biggl[
({\bf e}^\theta \otimes {\bf e}^\theta ){\partial ^2Y_{jm}\over \partial \theta
^2}
+({\bf e}^\varphi \otimes {\bf e}^\varphi )\left(
{\partial ^2Y_{jm}\over \partial \varphi ^2}+\sin \theta \cos \theta
{\partial Y_{jm}\over \partial \theta }\right) \cr
&~~~~~~~~~~~~~~~~
+({\bf e}^\theta \otimes {\bf e}^\varphi +{\bf e}^\varphi \otimes {\bf
e}^\theta )
\left( {\partial ^2Y_{jm}\over \partial \theta \partial \varphi }
-\cot \theta {\partial Y_{jm}\over \partial \varphi }\right)
\biggr] .\cr
}\eqn\badf$$

To solve for $F_j(\xi;\zeta)$,
we take the $({\bf e}^\xi \otimes {\bf e}^\xi )$ component of the
Laplacian acting on
eqn.~\bad , and after applying the
substitutions in eqn.~\badzc,
the coefficient of ${\bf e}^\xi \otimes {\bf e}^\xi$
term in \bac ~ becomes
$${\partial ^2F_j(\xz )\over \partial \xi ^2}+6\coth [\xi ]
{\partial F_j(\xz )\over \partial \xi }+
\left[ (\zeta ^2+3)+6\coth ^2[\xi ]
-{j(j+1)\over \sinh ^2[\xi ]}
\right] F_j(\xz  )=0.\eqn\badzd$$
With the change of variable $\phi_j(\xz  )=\sinh ^2[\xi ]\cdot F_j(\xz  ),$
one recovers the differential equation for the {\it scalar}
hyperbolic radial functions
(see, for example, [\bgt] eqn. (5.21)):
$${\partial ^2\phi_j(\xz ) \over \partial \xi ^2}+2\coth [\xi ]
{\partial \phi_j(\xz ) \over \partial \xi }+
\left[ (\zeta ^2+1)-{j(j+1)\over \sinh ^2[\xi ]}
\right] \phi _j(\xz  )=0. \eqn\scaleq$$
Consequently,
$$F_j(\xi ;\zeta )=N_j(\zeta )\cdot
\sinh ^{j-2}[\xi ]~~{~~~d^{j+1}\over d(\cosh [\xi ])^{j+1}}\cos [\zeta \xi ]
\eqn\badze$$
where the normalization
$N_j(\zeta )$ is determined in the
following, and we have imposed regularity at the origin.

For future reference, for $j=2$
$$F_2(\xi ;\zeta )={N_2(\zeta)\over \sinh ^3[\xi ]}\cdot
\Bigl[ 3\zeta ^2\coth [\xi ]\cos [\zeta \xi ]
+(\zeta ^3+\zeta -3\zeta \coth ^2[\xi ])\sin [\zeta \xi ]
\Bigr] .\eqn\badzf$$
and
from eqn.~\badzc
$$\eqalign{
G_2(\xi ;\zeta )&=N_2(\zeta )\cdot
\sinh [\xi ]\cdot \Biggl[ \Big\{ {\zeta ^2\coth [\xi ]\over 4}~
\Bigl( 1-2\zeta ^2-3\coth ^2[\xi ]\Bigr) \Bigr\} \sin [\zeta \xi ] \cr
&~~~~+\Bigr\{ {-\zeta ^5-2\zeta ^3-\zeta \over 4}+
{\zeta \coth ^2[\xi ]\over 4}\Bigl( \zeta ^2-2+3\coth ^2[\xi ]\Bigr)
\Bigl\} \cos [\zeta \xi ]\Biggr] ,\cr
H_2(\xi ;\zeta )&=N_2(\zeta )\cdot {1\over \sinh [\xi ]}\cdot \Biggl[ \Bigl\{
{\zeta ^2\over 6}\Bigl( \zeta ^2+4-6\coth ^2[\xi ]\Bigr) \Bigr\}
\cos [\zeta \xi ] \cr
&~~~~~~~~~~+ \Bigl\{
{\zeta \coth [\xi ]\over 2}\Bigl( 2\coth ^2[\xi ]-\zeta ^2-2\Bigr) \Bigr\}
\sin [\zeta \xi ]\Biggr] ,\cr
I_2(\xi ;\zeta )&=N_2(\zeta )\cdot \sinh [\xi ]\cdot \Biggl[
\Bigl\{ {\zeta ^2\coth [\xi ]\over 12}\Bigl(-5-2\zeta ^2+3\coth ^2[\xi ]
\Bigr) \Bigl\} \cos [\zeta \xi ]\cr
&~~+
\Bigl\{ {-\zeta ^5-4\zeta ^3-3\zeta \over 12}+{\zeta \coth ^2[\xi ]\over 4}
\Bigl( -\coth ^2[\xi ]+\zeta ^2+2\Bigr) \Bigr\} \sin [\zeta \xi ]\Biggr] .\cr }
\eqn\tta$$

Similarly, for the magnetic parity, the tensor field
must take the form
$$\eqalign{
{\bf T}^{B, jm}(\xi ,\theta ,\varphi ; \zeta )&=~~
U(\xi ; \zeta )~(
{\bf e}^{\tilde a}\otimes {\bf e}^{\xi }
+{\bf e}^{\xi }\otimes {\bf e}^{\tilde a})~
{\bf L}_{\tilde a}Y_{jm}(\theta ,\varphi ) \cr
&~~+V(\xi ; \zeta )~
({\bf e}^{\tilde a}\otimes {\bf e}^{\tilde b}+
{\bf e}^{\tilde b}\otimes {\bf e}^{\tilde a})~
{\bf L}_{\tilde a}\tilde \nabla _{\tilde b}
Y_{jm}(\theta ,\varphi ).\cr }
\eqn\bae$$
The magnetic parity modes do not
contribute to the CMB anisotropy because their component
along the $({\bf e}^\xi \otimes {\bf e}^\xi )$ direction vanishes;
therefore, we do not give their explicit form.

\noindent
{\bf Normalization of the Hyperbolic Tensor Harmonics.}
To normalize the tensor harmonics,
we impose the condition
$$\eqalign{\int _0^\infty d\xi &\int d\Omega \sqrt{\hat{g}_{(3)}}~
T_{ij}({\bf x};\zeta , P, j, m)^*~~ T^{ij}({\bf x};\zeta ', P', j', m')\cr
&
= \delta (\zeta -\zeta ')
\cdot \delta _{P, P'} \cdot \delta _{j, j'} \cdot \delta _{m, m'} \cr }
\eqn\zxa $$
where $P$ indicates mode parity and  ${\bf x}= (\xi ,\theta ,\phi ).$

Because the tensor harmonics are eigenfunctions of a self-adjoint operator,
the inner product is
proportional to a $\delta $-function. Thus $N_j(\zeta)$ is determined by
the coefficients of $e^{\pm i \zeta \xi}$ in the asymptotic expansion
of the tensor harmonics for large $\xi $.
(For $\xi \rightarrow 0$, $F_j, H_j,G_j, I_j \rightarrow 0$.)
In comparing the asymptotic behaviors of $F$, $G,$ $H$ and $I,$
it is more meaningful to consider the rescaled
quantities $\hat F_j(\xz  )=F_j(\xz  ),$ $\hat G_j(\xz  )=G_j(\xz  )/
\sinh ^2[\xi ],$ $\hat H_j(\xz  )=H_j(\xz  )/\sinh [\xi ],$
and $\hat I_j(\xz  )=I_j(\xz  )/\sinh ^2[\xi ]$,
components with respect to a normalized `vielbein' basis.
{}From eqn.~\badzc , $\hat F\sim \sinh ^{-3}[\xi ],$
$\hat H\sim \sinh ^{-2}[\xi ],$ and
$\hat G, ~\hat I\sim \sinh ^{-1}[\xi ].$
For $\xi \gg \lambda$, $\hat G$ and $\hat I$ dominate,
exactly as one would expect. That is, at large distances
a spherical gravitational wave should locally resemble a plane
gravitational wave propagating in the radial direction.

To compute the coefficient of the $\delta $-function in eqn.~\zxa ,
we impose a boundary condition at $\xi =\xi _{max}$ (for specificity
say Dirichlet boundary conditions) and take the limit $\xi _{max}\to \infty .$
For $\xi _{max}\gg 1,$
the integral in eqn.~\zxa ~is dominated by the
$G$ and $I$ components.  These may be approximated by
their large-$\xi $ asymptotic forms, starting
with eqn.~\badze ~and  substituting
$\sinh [\xi ]\to (e^\xi /2)=(w/2).$  This gives
$$ \eqalign{
F_j(\xi ;\zeta )&\approx  4~N_j(\zeta )~w^{-3}
\left( {~d\over dw}\right) ^{j+1}~
[ w^{+i\zeta }+ w^{-i\zeta }] \cr
&=4~N_j(\zeta )~e^{-3\xi } \cdot
\Bigl[ (i\zeta )_j~e^{+i\zeta \xi }+c.c.~\Bigr] .}
\eqn\zxb $$
Here $(x)_j$ is shorthand for $x(x-1)\ldots (x-j).$
Using eqn.~\badzc , we obtain
$$\eqalign{
H_j(\xi ;\zeta )&\approx {4~N_j(\zeta )~e^{-\xi }\over
j(j+1)} \cdot
\Bigl[ (i\zeta )\cdot (i\zeta )_j~e^{+i\zeta \xi }+c.c.~\Bigr] ,\cr
I_j(\xi ;\zeta )&\approx {2~N_j(\zeta )~e^{+\xi }\over
j(j+1)(j-1)(j+2)} \cdot
\Bigl[ (i\zeta +1)\cdot (i\zeta )
\cdot (i\zeta )_j~e^{+i\zeta \xi}+c.c.~\Bigr] ,\cr
G_j(\xi ;\zeta )&\approx {N_j(\zeta )~e^{+\xi }\over
(j-1)(j+2)} \cdot
\Bigl[ (i\zeta +1)\cdot
(i\zeta )\cdot (i\zeta )_j~e^{+i\zeta \xi }+c.c.~\Bigr] .\cr
}\eqn\zxc$$
Consequently, for $(\xi \gg 1),$
$$\eqalign{{\bf T}({\bf x};\zeta , E, j, m)\approx &
\Biggl[ { 2N_j(\zeta )e^{\xi}\over j(j+1)(j-1)(j+2)}\cdot
(i\zeta +1)\cdot
(i\zeta )\cdot
(i\zeta )_j\cdot
e^{+i\zeta  \xi}\cr
&~~~~~\times \Bigl[ \nabla _{\tilde a}\nabla _{\tilde b}Y_{jm}(\theta , \phi )
+\delta _{\tilde a\tilde b}{1\over 2}j(j+1)Y_{jm}(\theta , \phi )
\Bigr]
\Bigl( \hat {\bf e}^{\tilde a}\otimes \hat {\bf e}^{\tilde b} \Bigr)
\Biggr] \cr
&+c.c. \cr
}\eqn\xzd$$
Inserting this asymptotic expression
into eqn.~\zxa ~and factoring out
$\delta _{j j^\prime} ~\delta_{m m^\prime}$ gives
$$\eqalign{
\int _0^{\xi _{max}}&d\xi ~\sinh ^2[\xi ]\int d\Omega ~
T_{ij}({\bf x};\zeta , E, j, m)
T^{ij}({\bf x};\zeta , E, j, m)^*~\cr
\approx & \vert {N_j} (\zeta ) \vert^2\cdot {4
\zeta ^2(\zeta ^2+1)~[\zeta ^2(\zeta ^2+1^2)\ldots (\zeta ^2+j^2)]
\over j^2(j+1)^2(j-1)^2(j+2)^2}\cr
&\times{\xi _{max}\over 2}\int d\Omega \left|
 \nabla _{\tilde a}\nabla _{\tilde b}Y_{jm}
+{1\over 2}\delta _{\tilde a\tilde b}j(j+1)Y_{jm} \right| ^2.\cr
}\eqn\xzf$$

For the angular integration, note that
$$\eqalign{
\int d\Omega \Bigl( \nabla _{\tilde a}\nabla _{\tilde b}Y_{jm}\Bigr) ^*~
\nabla ^{\tilde a}\nabla ^{\tilde b}Y_{jm}
&=-\int d\Omega \Bigl( \nabla ^{\tilde a}\nabla _{\tilde a}\nabla _{\tilde b}
Y_{jm}\Bigr) ^*~ \Bigl( \nabla ^{\tilde b}Y_{jm}\Bigr) \cr
&=-\int d\Omega \Bigl( \nabla _{\tilde a}\nabla _{\tilde b}\nabla ^{\tilde a}
Y_{jm}\Bigr) ^*~\Bigl(  \nabla ^{\tilde b}Y_{jm}\Bigr) \cr
&=-\int d\Omega \Bigl( \nabla _{\tilde b}\Bigl\{ \nabla ^2 +{1\over 2}R\Bigr\}
Y_{jm}\Bigr) ^*~\Bigl(  \nabla ^{\tilde b}Y_{jm}\Bigr) \cr
&=~~[~j(j+1)~]\cdot [j(j+1)-1~] \; ,\cr
}\eqn\xze$$
given that
$g^{\tilde a\tilde c}~[\nabla _{\tilde a},\nabla _{\tilde b}]~\nabla _{\tilde
c}f =
g^{\tilde a\tilde c}
{R_{\tilde a\tilde b\tilde c}}^{\tilde d}\nabla _{d}f ={R_{\tilde b}}^{\tilde
d}
(\nabla _{\tilde d}f)={1\over 2}R~\nabla _{\tilde b}f.$
Since the unit two-sphere is isotropic,
$R_{\tilde a\tilde b}={1\over 2}\delta _{\tilde a\tilde b}R$.
Also, for the two-sphere $R=2.$
Thus the integral on the last line of eqn.~\xzf ~is equal to
$[j(j+1)][{1\over 2}j(j+1)-1],$ and
eqn.~\xzf ~reduces to
$$ \vert {N_j} (\zeta )\vert^2\cdot
{4[j(j+1)][{1\over 2}j(j+1)-1]\over j^2(j+1)^2(j-1)^2(j+2)^2}
\times \zeta ^2(\zeta ^2+1)[\zeta ^2(\zeta ^2+1^2)\ldots (\zeta ^2+j^2)]
\times{\xi _{max}\over 2},\eqn\xzg$$
from which it follows that
$$N_j(\zeta )={1\over \sqrt{\pi }}~
{\sqrt{ j(j+1)(j-1)(j+2)}
\over
 \zeta \sqrt{1 + \zeta^2}
\sqrt{\zeta ^2(\zeta ^2+1^2)\ldots (\zeta ^2+j^2)}} \; .\eqn\xzh$$
The $\pi ^{-1/2}$ comes from passing to the continuum
and the phase of $N_j(\zeta )$ has been chosen to be real.

\noindent
{\bf Flat Tensor Harmonics.}
The general form of the flat modes is needed to identify the
Bunch-Davies vacuum in the next section.
These may be
regarded as the large-$\zeta $, small-$\xi $ limit
of the hyperbolic modes, a limit in which the effects of spatial
curvature disappear.
We use lower case letters to denote the flat
analogues of hyperbolic quantities.
In particular,
$$f(r,\omega )=\bar n_j(\omega )\cdot r^{j-2}\left( {1\over r}
{~d\over dr}\right) ^{j+1}\cos [\omega r].\eqn\bafa$$
Similarly, eqn.~\badzc ~ is modified to
$$\eqalign{
h_j(r; \omega )&={r^2\over j(j+1)}\left[ {\partial f_j\over \partial r}
+{3\over r}f_j\right] ,\cr
i_j(r; \omega )&={r^2\over (j+2)(j-1)}\left[ 2{\partial h_j\over \partial r}
+{4\over r}h_j-f_j\right] ,\cr
g_j(r; \omega )&={1\over 2}\bigl[ j(j+1)i_j(r)-r^2f_j\bigr] .\cr
}\eqn\ttb$$
In particular, for $j=2$
$$\eqalign{
f_2(r;\omega )&=
\left( {3\omega ^2\over r^3}\right) \cos [\omega r]
+\left( {-3\omega \over r^4}+{\omega ^3\over r^2}\right) \sin [\omega r],
\cr
g_2(r;\omega )&=
\left( {-3\omega ^2\over r}\right) \cos [\omega r]
+\left( {-\omega ^5r^2\over 4}-\omega ^3+{3\omega \over r^2}
\right) \sin [\omega r],
\cr
h_2(r;\omega )&=
\left( {-\omega ^2\over 2r^2}+{\omega ^4\over 6}\right) \cos [\omega r]
+\left( {-\omega ^3\over 3r}+{\omega \over 2r^3}\right) \sin [\omega r],
\cr
i_2(r;\omega )&=
\left( {-\omega ^2\over 2r}\right) \cos [\omega r]
+\left( {-\omega ^5r^2\over 12}-{\omega ^3\over 6}+{\omega \over 2r^2}
\right) \sin [\omega r].
\cr }\eqn\ttc$$

In order to calculate the normalization, one needs the asymptotic
behavior for large $r$:
$$\eqalign{
f_j(\xi ;\omega )&\approx \bar n_j(\zeta )\cdot {1\over 2r^3}
\Bigr[ ~(i\omega)^{j+1}~e^{+i\omega r}+c.c.~\Bigr] ,\cr
h_j(\xi ;\omega )&\approx \bar n_j(\zeta )\cdot {1\over j(j+1)}\cdot {1\over
2r}
\Bigr[ ~(i\omega)^{j+2}~e^{+i\omega r}+c.c.~\Bigr] ,\cr
i_j(\xi ;\omega )&\approx \bar n_j(\zeta )\cdot
{1\over j(j+1)(j-1)(j+2)}\cdot r \cdot
\Bigr[ ~(i\omega)^{j+3}~e^{+i\omega r}+c.c.~\Bigr] ,\cr
g_j(\xi ;\omega )
&\approx \bar n_j(\zeta )\cdot {1\over 2(j-1)(j+2)}\cdot r\cdot
\Bigr[ ~(i\omega)^{j+3}~e^{+i\omega r}+c.c.~\Bigr] .\cr
}\eqn\ttd$$
Again $i_j$ and
$g_j$ dominate, and following the same steps as
for the hyperbolic modes we get
$$\bar n_j(\omega )=
{\sqrt{j(j+1)(j+2)(j-1)}
\over \sqrt{\pi} \omega ^{j+3}} \; .
\eqn\ttf$$
This is the $\zeta \rightarrow \omega \gg j$ limit of
eqn. \xzh .

\noindent
{\bf Normalization of the Time Dependent Mode Functions.}
We may define the antisymmetric bilinear form
$$\langle {\cal U}, {\cal V}\rangle =
-\int _\Sigma d\Sigma ^\mu ~{\cal U}_{\alpha \beta }(\Sigma )~
(i\overleftrightarrow D_\mu ){\cal V}^{\alpha \beta }(\Sigma ),
\eqn\gba$$
where ${\cal U}, {\cal V}$ are solutions to eqn.~\xxa . This
product is analogous to the Klein-Gordon product for scalar
field modes.
Eqn.~\xxa ~insures that
the current ${\cal U}_{AB}(\Sigma )~
(i\overleftrightarrow \nabla_\mu ){\cal V}^{AB}(\Sigma )$ is
conserved and thus that  $\langle {\cal U}, {\cal V}\rangle $
is invariant under deformations of the surface $\Sigma .$
In order that the modes for a spacetime $M$ orthonormalized
with respect to \gba ~are associated with operators that
satisfy the customary canonical commutation relations, it is necessary
to choose $\Sigma $ in \gba ~so that $\Sigma $ is a Cauchy surface
for $M.$ The product $\langle ~~,~~\rangle $ is the same
for all Cauchy surfaces for the spacetime $M.$
A Cauchy surface is a spacelike hypersurface which each non-spacelike
curve intersects once and only once [\hawking]. In this paper
we shall consider both the case where $M$ is just region I,
in which case a surface of constant region I hyperbolic time
serves as a convenient choice of Cauchy surface, and also
the case where $M$ is all of maximally extended de Sitter
space, in which case the surface defined by $\tau =0$ in the
region II hyperbolic coordinates serves as a convenient
Cauchy surface.

For the hyperbolic modes defined in eqn.~\lyaa ~ in region I one
calculates
$$\eqalign{
&\langle {\cal T}(\zeta ,P,j,m), {\cal T}(\zeta ',P',j',m')^*\rangle \cr
&=-\int _\Sigma d^3x~a^3(\eta )
\biggl[
  {\cal T}_\alpha ^\beta  ({\bf x}, \eta ;\zeta  , P, j, m )
  (i\overleftrightarrow D_{\hat 0})~
  {\cal T}^\alpha _\beta  ({\bf x}, \eta ;\zeta ', P', j', m' )^*~
\biggr] \cr
&=n^2 (\zeta )~
2~\zeta ~(\zeta ^2+1)\cdot \delta (\zeta -\zeta ')\cdot \delta _{P, P'}
\cdot \delta _{j, j'}\cdot \delta _{m, m'} \cr }
\eqn\fhb$$
where $D_{\hat 0}=(1/a)D_\eta =-\sinh [\eta ]D_\eta .$
Note that eqn.~\fhb ~is independent of $\eta .$ This may be seen by
applying $D_\eta $ to eqn.~\fhb ~and using eqn.~\jjc .
For the mixed tensor representation chosen above,
the covariant derivative $D_{\hat 0}$ may
be replaced with the ordinary derivative $\partial _{\hat 0}.$
So, reading off,
$$n(\zeta )={1 \over\sqrt{2\zeta(\zeta^2+1)}}.
\eqn\lyz$$

As the
initial conditions are determined by the bubble which extends outside
of region I, a proper Cauchy surface for initial conditions for
an open universe
extends outside of region I as well.
It was shown in [\norm] that for some cases,
inner products taken on a Cauchy surface in region II
agreed with those on fixed time surfaces in region I and V,
even though the latter do not make up a Cauchy surface for
the whole spacetime.  The norms agreed for scalar fields
with sufficiently fast falloff at infinity, a condition
satisfied by subcurvature modes of the Laplacian
(modes with eigenvalue $\zeta^2 +1 \ge 1$).
There are some differences in extending this comparison
between norms taken in I and V and norms taken in II for gravity
waves. When the gravity waves are continued across the light cone,
into region II, time and
space are interchanged and the gauge choice becomes
$h_{u \mu} = 0$ instead of $h_{\eta \mu} = 0$.
As $u$ is a spacelike coordinate, this is not the usual gauge
for metric perturbations.
It may be thought of
of as analogous to axial gauge (where $A_3 = 0$ ) rather than
Coulomb gauge (where $A_0 = 0$) in electrodynamics.
Secondly, the inner product in region II
involves a wronskian in (the analytic continuation of) $\xi$,
and is more complicated due to the tensor structure in $\xi$.
As shown in appendix B, this inner product on a Cauchy surface
in region II
coincides with the inner product
in regions I, V, for modes which have sufficiently fast falloff at
infinity.
The fields with sufficiently
fast falloff for gravity waves are again subcurvature modes,
$0 \le \zeta^2 < \infty$.
As the falloff for the gravity wave modes is as fast
as that for the scalars in regions I, V (both go as
$\sinh^{-1} [\xi ]$ for large $\xi$), it was reasonable to expect this.

\chapter{Identifying the Gravitational Wave Bunch-Davies Vacuum}

The preceding section gave the mode expansion for the
linearized gravitational waves in region I hyperbolic coordinates,
which are the natural coordinates for the open universe inside
the  bubble.  In this section the initial Bunch-Davies
vacuum is expressed in terms of these open hyperbolic modes.
In the open universe inflationary
scenario,
the Bunch-Davies vacuum is a preferred quantum state for de Sitter
space in the sense that it is a weak attractor:
any initial quantum state for perturbations from de Sitter space,
subject only to the requirement that the initial energy density be
finite, approaches the Bunch-Davies vacuum to arbitrary accuracy
after a sufficient amount of inflationary expansion. The convergence
is weak rather than strong because the initial perturbations
are not erased but rather pushed to larger and larger scales,
so that for an observer able to probe only a fixed physical volume,
the perturbations seem to disappear.

The Bunch-Davies vacuum is physically characterized using the flat
coordinates for de Sitter space.  In these
coordinates, at sufficiently early times,
a mode of fixed co-moving wavenumber evolves as if it were a mode
in Minkowski space.
For the flat coordinates the line element is
$$ ds^2=-dt_f^2+e^{2t_f}\cdot \Bigl[ dr_f^2+r_f^2~d\Omega _{(2)}^2\Bigr]
\eqn\pba $$
where $-\infty <t_f<+\infty ,$ or in terms of flat
conformal time $\eta _f=-e^{-t_f}$
$$ ds^2={1\over \eta _f^2}\cdot \Bigl[
-d\eta _f^2+dr_f^2+r_f^2~d\Omega _{(2)}^2\Bigr] .
\eqn\pbb $$

Early on, when the mode is subhorizon, so that
there are many oscillations within an expansion time, one identifies
the mode with positive frequency asymptotic behavior with an annihilation
operator of the Bunch-Davies vacuum. These are the modes that
behave as ${e^{-ik\eta_f} /\eta_f }$ for $\eta_f \to -\infty ,$
where $\eta_f $ is flat conformal time.

\vskip -.5in
\def\Three{3}
\LoadFigure\Three{\baselineskip 13 pt
\noindent\narrower\ninerm
{\ninebf ---Flat Coordinates for de Sitter Space.}
The region covered by the flat coordinates is shown in
a conformal diagram for all of maximally extended de Sitter
space, identical to that in Fig.~2. Although the
flat coordinates cover only half of maximally extended de Sitter space,
the diagonal line, which is the boundary of the region covered
by the flat coordinates, represents an
initial value surface for all of de Sitter space.}
{\epsfysize 4.5truein}{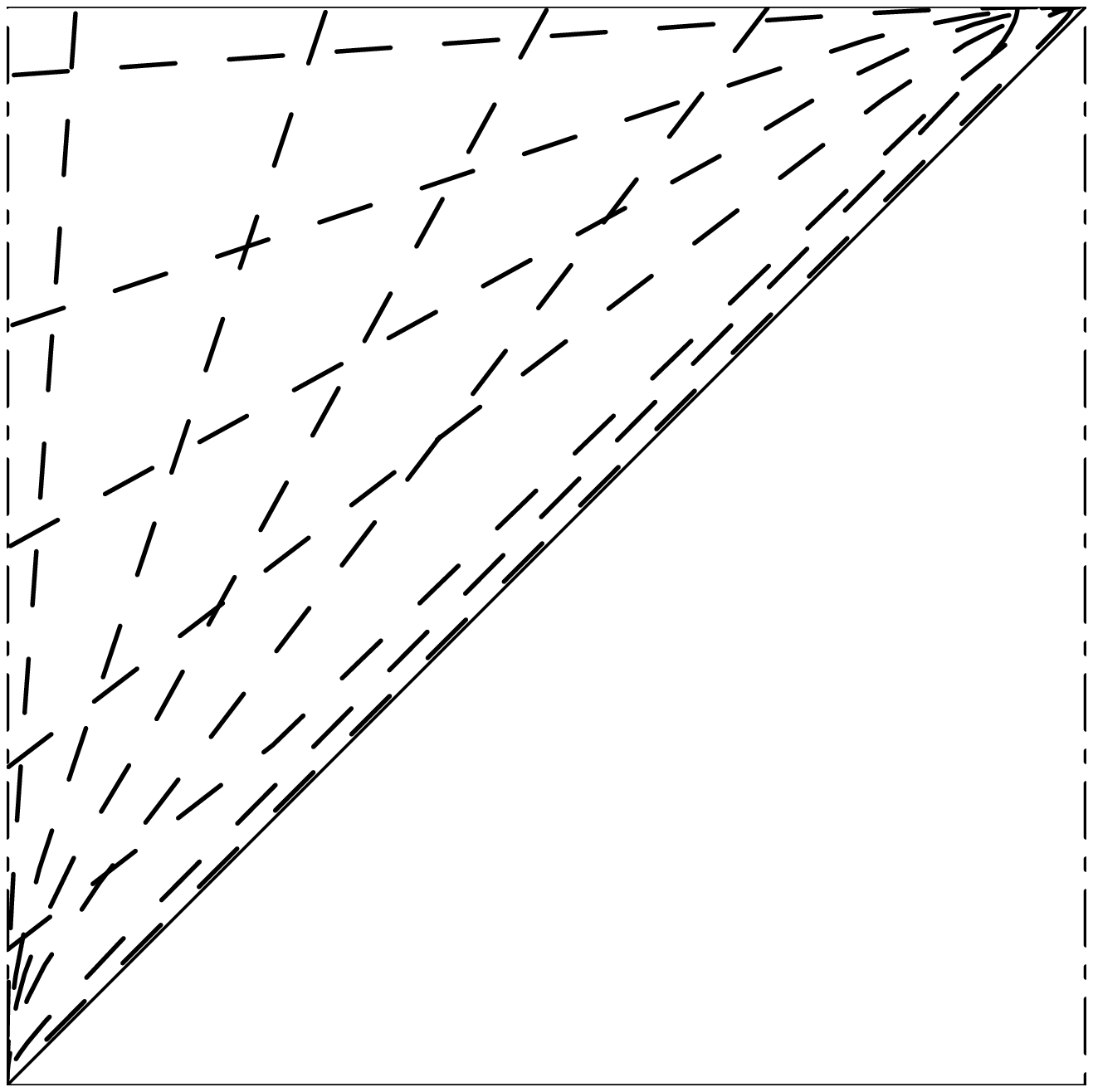}{}

To relate the flat and hyperbolic coordinates one can
embed de Sitter space in (4+1)-dimensional
Minkowski space as described in appendix A.
Using $\sinh [t_h]=-1/\sinh [\eta ],$ $\cosh [t_h]=-\coth [\eta ],$
and $\coth [t_h]=\cosh [\eta ]$, one gets that
the relation between the coordinate systems is
$$ \eqalign{
r_f&={\sinh [\xi ]\over \cosh [\xi ]+\cosh [\eta ]},\cr
\eta _f&={\sinh [\eta ]\over \cosh [\eta ]+\cosh [\xi ]}.\cr
}\eqn\pbf$$
The flat coordinates do not cover all of de Sitter space but only
half of maximally extended de Sitter space, as indicated in the
conformal diagram in Fig.~3. Nevertheless the flat coordinates
cover enough of de Sitter space to contain an
initial value surface for all of de Sitter space.
The null surface indicated by the dashed
diagonal line, the boundary of the
region covered by the flat coordinates, is such a
surface.\footnote\dagger{Technically, because this
surface is null it is not a Cauchy surface, but
initial data on this surface can be continued everywhere in de Sitter
space.}

To identify the hyperbolic modes of {\it positive} frequency with
respect to the Bunch-Davies vacuum, it is convenient to use the
null coordinates in the flat chart
$$\eqalign{
u_f&=\eta_f +r_f,\cr
v_f&=\eta_f -r_f \; . \cr }\eqn\iba$$
One has $v_f<0$ and $\vert u_f \vert<-v_f.$
We also define the region I hyperbolic null coordinates
$$\eqalign{
U&=\eta +\xi ,\cr
V&=\eta -\xi \cr }\eqn\ibb$$
where $V<0$ and $\vert U \vert < -V.$
Using the above relations between $(\eta,\xi)$ and
$(\eta_f, r_f)$ one can see that
two sets of null coordinates are
related according to
$$\eqalign{
U&=\ln \left[ {1+u_f\over 1-u_f} \right] ,\quad
V=\ln \left[ {1+v_f\over 1-v_f} \right] \cr }.\eqn\ibc$$
Conversely,
$u_f = \tanh {U\over 2} $ and $
v_f = \tanh {V\over 2}\; . $
Thus we see that the region I hyperbolic coordinates only cover the
region $-1<v_f<0,$ $\vert u_f\vert <-v_f$
in terms of the flat coordinates.

The positive frequency modes with respect to open hyperbolic light
cone coordinates in region I are
$$\eqalign{
e^{-i\zeta U}&=\left[ {1+u_f\over 1-u_f} \right] ^{-i\zeta },\quad
e^{-i\zeta V}=\left[ {1+v_f\over 1-v_f} \right] ^{-i\zeta }.\cr }\eqn\ibd$$
However, a full mode function for the Bunch-Davies vacuum
must be specified in both regions I and V.
A region I mode function must be continued into region $V$ in order to
calculate its inner product on the full space and
the choice of analytic continuation distinguishes between positive and
negative frequency Bunch-Davies modes.
The identification of
the positive frequency modes for the Bunch-Davies vacuum is determined
by the analytic properties of the
factors in $e^{-i\zeta U}$ and $e^{-i\zeta V}$
appearing in the hyperbolic mode functions above.  The
other factors in the hyperbolic mode functions
contain only isolated singularities and no branch cuts, and thus will
be seen to be irrelevant in making this identification.
The bubble interior lies within the strip $-1<u_f<+1.$ Outside
this strip one encounters a branch cut, taken here to
lie near the real axis starting at $u_f=1,$ passing through $+\infty ,$
coming back from $-\infty ,$ and finally ending at $u_f=-1,$ as indicated in
Fig.~4.
Thus $[(1+u_f)/(1-u_f)]^{i\zeta }$ has two possible
continuations to the entire real line---one through the upper half-plane
above the branch cut, and another through the lower half-plane
below the branch cut.

\vskip -.5in
\def\Four{4}
\LoadFigure\Four{\baselineskip 13 pt
\noindent\narrower\ninerm
{\ninebf ---Analytic Continuation of the Mode Functions.}
The points $u_f=-1$ and $u_f=+1$ are branch points. To impose
single-valuedness, we
take a branch cut to start $u_f=+1,$ extend to $u_f=+\infty ,$
come back from $u_f=+\infty ,$ and finally end at $u_f=-1.$
The analytic continuation through the upper half-plane onto
the remainder of the real line corresponds to
the Bunch-Davies positive frequency modes. Similarly, the
continuation through the lower half-plane corresponds to
the Bunch-Davies negative frequency modes. }
{\epsfysize 3truein}{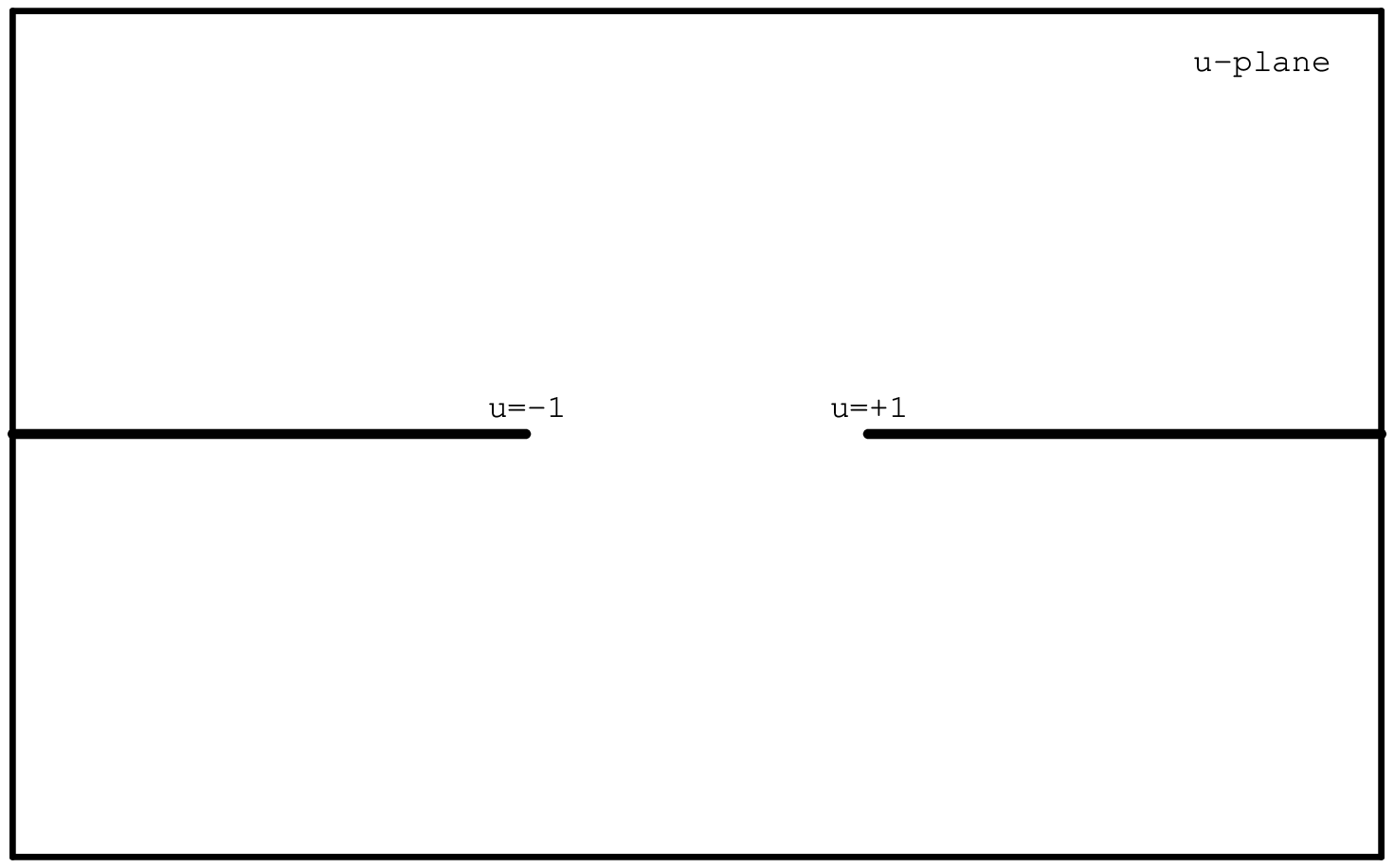}{}

One can identify the positive frequency Bunch
Davies modes by requiring they have vanishing overlap with
the flat negative frequency
modes $e^{i \omega u_f}$, ($\omega >0$).  Thus one requires
$$\int _{-\infty }^{+\infty }du_f~ e^{i\omega u_f}~
\left( {1+u_f\over 1-u_f}\right) ^{i\zeta }=0\eqn\ibf$$
for all $\omega >0.$  This is satisfied by the analytic continuation
through the upper half-plane above the branch cut,
as deforming the contour toward $+i\infty $ makes this integral vanish.
As a result, positive frequency Bunch Davies modes correspond to
analytic continuation through the upper half plane into region $V$.
This identification
generalizes to gravitational wave hyperbolic modes and
other more complicated mode functions of the form
$$R(u_f,v_f)~\left( {1+u_f\over 1-u_f}\right) ^{i\zeta }+
S(u_f,v_f)~\left( {1+v_f\over 1-v_f}\right) ^{i\zeta }\eqn\ibg$$
where $R(u_f,v_f)$ and $S(u_f,v_f)$ are rational functions.

\vskip -.5in
\def\Five{5}
\LoadFigure\Five{\baselineskip 13 pt
\noindent\narrower\ninerm
{\ninebf ---Dual Pair of Flat Coordinates.}
Both panels are conformal diagrams for maximally extended de Sitter
space. The shaded triangle in Fig.~5(a) shows the region covered
by the flat null coordinates $(u_f,v_f),$ while in Fig.~5(b) the region
covered by null coordinates $(\tilde u_f,\tilde v_f),$ is shown.}
{\epsfysize 2.5truein}{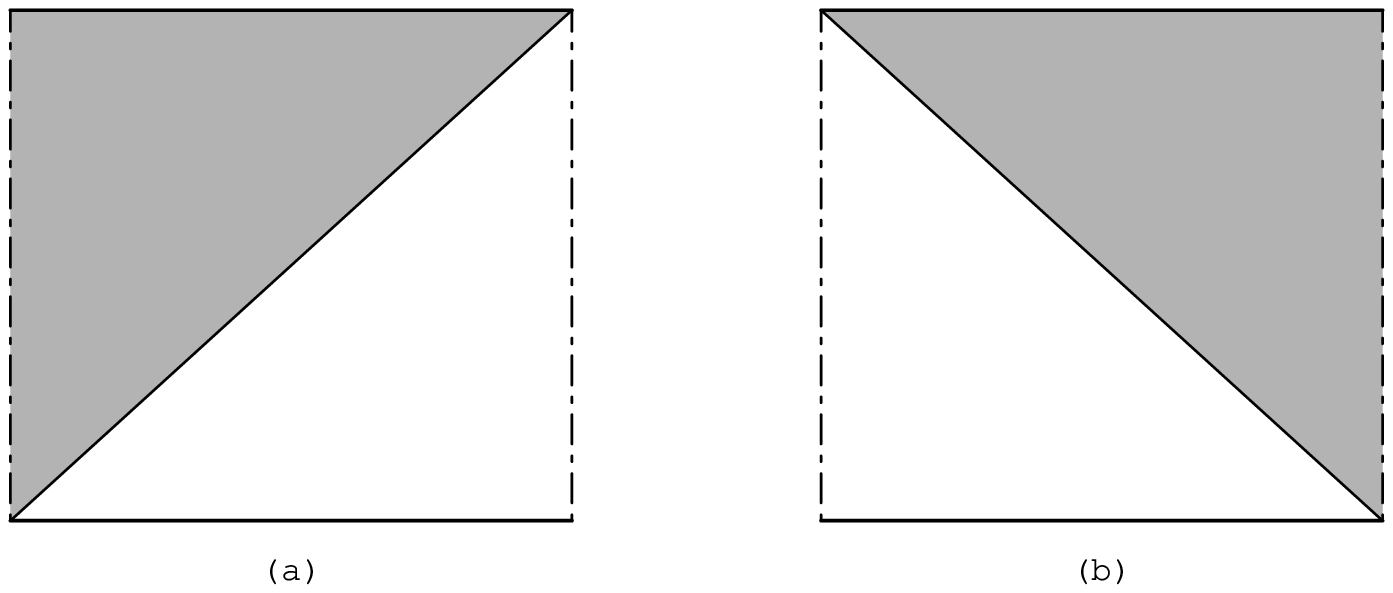}{}

We shall use two sets of flat coordinates
$(u_f,v_f)$ and $(\tilde u_f,\tilde v_f),$
as indicated in Fig.~5, related by a spatial
reflection symmetry of de Sitter space that maps region
I into region V. It may be shown that
$$\tilde u_f={-1\over v_f},\quad \tilde v_f={-1\over u_f},\eqn\fliprel$$
either directly by using closed coordinates or by the
analytic methods in appendix A.

This transformation relates the flat coordinates $(u_f,v_f)$ covering the
upper left triangular wedge of de Sitter space to
the coordinates $(\tilde u_f,\tilde v_f)$
covering the upper right triangular
wedge, as indicated in Fig.~5.
As discussed earlier, for positive
frequency mode functions the analytic continuation is with positive
imaginary part, $(-1) = e^{i\pi }$, and so
$$
\eqalign{
& \left( {1+u_f\over 1-u_f}\right) ^{-i\zeta }
=\left((-1)^{-1}{\tilde 1+v_f\over 1-\tilde v_f}\right) ^{i\zeta }
\equiv (-1)^{- i \zeta} e^{i \zeta \tilde{V}}\cr
&\left( {1+v_f\over 1-v_f}\right) ^{-i\zeta }
=\left((-1)^{-1} {\tilde 1+u_f\over 1-\tilde u_f}\right) ^{i\zeta }
\equiv  (-1)^{-i \zeta} e^{i \zeta \tilde{U}} \; .\cr
}\eqn\ibj$$
As a result, positive hyperbolic
frequency modes $e^{-i\zeta U}, e^{-i \zeta V}$
in region I  become negative hyperbolic frequency modes in
region V with an attentuation factor $e^{-\pi \zeta}$,
and negative frequency region I hyperbolic modes become
positive frequency hyperbolic modes in region V with an
amplification factor $e^{\pi \zeta}$.
The result of all this is that the positive frequency modes in
region I are the modes proportional to
$e^{i U \zeta}\, , \; e^{i V \zeta}$ for all values of $\zeta$
plus the continuation into region $V$ using the upper half plane.
The negative frequency modes are the same collection of modes with
the opposite analytic continuation into region V.

We may normalize the modes by combining the fixed time surfaces of
regions I and V.  As shown in [\norm] for scalar
subcurvature modes, and in the appendix for gravity waves, for the
 subcurvature
modes under consideration here, the inner product is equivalent to
that taken on a Cauchy surface for all of de Sitter space. Let
${\cal T}^{(I)}(\zeta ,P,j,m)$
denote the modes with positive
hyperbolic frequency in region I, as defined in eqn.~\lyaa ,
normalized using only the region I fixed time surface, and let
${\cal T}^{(V)}(\zeta ,P,j,m)$
be the analogously defined modes in region V. It follows that
in region I one has the mode expansion,
$$\eqalign{
\sum_{Pjm}\int _0^\infty
&{d\zeta ~\over \sqrt{ 2\zeta (\zeta^2 +1)}}~ \{
\Bigl[ {\bf T}^{Ejm}(\xi ,\theta ,\phi ,\eta ;\zeta )\Bigr] ^r_{~s}
\biggl[ {e^{+\pi \zeta /2}\over (e^{+\pi \zeta }-e^{-\pi \zeta })^{1/2}}~~
T_h(\eta ; \zeta)~\hat b_I(\zeta ,P,j,m) \cr
&~~~+{e^{-\pi \zeta /2}\over (e^{+\pi \zeta }-e^{-\pi \zeta })^{1/2}}~~
T_h(\eta; -\zeta) ~\hat b_V%
(\zeta ,P,j,m)\biggr]
+h.c.  \}\cr
}\eqn\gfa$$
and similarly in region V
$$\eqalign{\sum_{jm}\int _0^\infty
&{d\zeta ~\over \sqrt{ 2\zeta (\zeta^2 +1)}}~\{
\Bigl[ {\bf T}^{Ejm}(\xi ,\theta ,\phi ,\eta ;\zeta )\Bigr] ^r_{~s}
\biggl[ {e^{+\pi \zeta /2}\over (e^{+\pi \zeta }-e^{-\pi \zeta })^{1/2}}~~
T_h(\eta ; \zeta)~\hat b_V(\zeta ,P,j,m)\cr
&~~~+{e^{-\pi \zeta /2}\over (e^{+\pi \zeta }-e^{-\pi \zeta })^{1/2}}~~
T_h(\eta ; -\zeta) ~\hat b_I %
(\zeta ,P,j,m)\biggr]
+h.c. \} \cr }\eqn\gfaz$$
where the annihilation operators $\hat b_I(\zeta ,P,j,m)$
and $\hat b_V(\zeta ,P,j,m)$ annihilate the Bunch-Davies vacuum
for the graviton field and satisfy the usual commutation relations
$$\eqalign{&[\hat b_A(\zeta ,P,j,m),~ \hat b_B(\zeta ',P',j',m')]=0,\cr
&[\hat b_A^\dagger (\zeta ,P,j,m),~ \hat b_B^\dagger (\zeta ',P',j',m')]=0,\cr
&[\hat b_A(\zeta ,P,j,m),~ \hat b_B^\dagger (\zeta ',P',j',m')]=
\delta _{AB}~\delta (\zeta -\zeta ')~\delta _{PP'}~
\delta _{jj'}~\delta _{mm'}~\cr }\eqn\lzza$$
where $(A,B=I,V).$
In the Bunch-Davies vacuum we observe a doubling in the number of
modes.
This
reflects the existence of correlations between regions I and V---or,
alternatively, between the inside and
the outside of the bubble.

Another way of understanding this result for the
vacuum is to note that
the scalar mode functions and mixed tensor (with
one raised and one lowered index, ${\cal T}^A_B$) mode functions
considered here
have the same time dependence.
As the choice of vacuum
depends only on the time dependence of the mode functions,
the time dependence in the
Bunch-Davies vacuum also has the same form for
both cases.\foot{J.D.C. thanks
K. Schleich for discussions about this.}

For the scalars there is a discrete vacuum mode [\norm] and here one also
is normalizable, by the same arguments.

\chapter{Gravity Waves and a Bubble with Finite Energy Difference}

In this section we take into account the effects of the nonvanishing
size of the critical bubble and of the nonvanishing difference in
energy density across the bubble wall. In the presence of a bubble,
one starts with a Bunch-Davies vacuum with $H=H_F$ as an initial
condition outside the bubble and then continues these modes across
the bubble wall into the open universe, where we idealize the expansion
rate to be a constant with $H=H_T.$ Of course, in realistic single-bubble
inflationary models $H$ inside the bubble is neither constant, nor
does it correspond to the true vacuum as the subscript $T$ suggests.
However, in order to capture the qualitative consequences
of changing $H$ across the bubble wall without getting involved in
the messy details of specific models, we consider the idealization
of an infinitely thin wall, with constant $H=H_T$ inside, for
calculating the gravitational waves.
Whereas earlier in the paper we set $H$ to unity, from here on
factors of $H$ will be displayed
explicitly. Given that the gravity waves and the scalars have the same
time dependence, and this is the only place the dependence on the wall
appears, one can carry over the results for scalar fields.
In the case at hand, we have a massless field before and after
tunneling and a finite energy difference across the wall.
The specific relevant
results are from [\YTSnew, \yssimple, \jdc] and are briefly
sketched below.

As will be
shown below, any bubble with $H_F \ne H_T$ has a finite radius and
extends outside the light cone (region I) covered by
the $(\eta, \xi,\theta,\phi)$ coordinate system.
By considering
a curve of constant field value outside the light cone
in Fig. 1a, one sees that the matching across the wall
thus has to be done in region II and its
Euclidean continuation.
The continuation into region II from region I is
$e^\eta \rightarrow -i e^{-u},
\; \xi \rightarrow \tau + i \pi/2$, and
so the metric in region II becomes
$$ds^2 = a^2(u) \cdot \biggl[
du^2 - d \tau^2 + \cosh^2 \tau d \Omega_{(2)}^2\biggr] \; .
\eqn\regtwometric$$
For an open universe $a(u) = 1/(H \cosh [u])$.
The analytic continuation of the basis functions is given in
somewhat more detail in appendix $B$ and is straightforward.
With the Euclidean continuation appropriate for
tunneling, one also, before nucleation, rotates the
space to Euclidean time $\tau \rightarrow i \tau_E$.

The wall has the same
description in the Euclidean and Lorentzian regions in
spacetime,
since it only depends on $u$ and not $\tau,\tau_E$.
For a thin wall,
the $u$ dependent scale factor is [\sasakinew]
$$a(u) = \cases{
{1 \over H_F \cosh [u]}, & for $u< u_R,$\cr
{1 \over H_T \cosh [u+\delta ]}, & for $u >u_R,$ \cr
}\eqn\inst $$
where
$${1\over H_T \cosh[u+\delta]}
={1 \over H_F\left[ \cosh[u-u_R] \cosh u_R
+ \sinh[u-u_R]\sqrt{\cosh^2[u_R] - (H_T/H_F)^2}\right]
}.\eqn\fna$$

\noindent
The metrics are explicitly matched at the same value of $a(u)$, the
bubble radius
$$R = {1 \over H_F \cosh [u_R]} 
= {1 \over H_T \cosh [u_R + \delta ]} \; .
\eqn\radius$$

The mode functions in this background have only their $u$-dependence
changing across the wall.
Requiring continuity in the unchanged $(\tau,\theta,\phi)$ dependence
forces the exterior mode functions to match onto interior modes
with the same value of $\zeta^2$.
To match the mode functions across this wall, we start with the inital
$u$ dependence corresponding to the analytic continuation of
the time $(\eta)$ dependence found in eqn.~\jje.  As
the exterior of the bubble has $H= H_F$, the false
vacuum mode functions
$e^{\pi \zeta/2} T_h(\eta \; \zeta)$, extend to region II
as
$$A_\zeta(u) =H_F ~\cosh [u]~ e^{i\zeta u}~{\biggl( \tanh [u] - i\zeta\biggr) }
\eqn\fnb$$
up to a factor of $i$.
These are matched across the wall onto modes of the
{\it true} vacuum (as there is some slow roll after tunneling this is
not exactly the true vacuum but is close enough for our purposes).
The corresponding
mode functions interior to the bubble, with $u > u_R$,
$H= H_T$, are
$$B_\zeta(u+ \delta)= H_T \cosh [u + \delta ]~ e^{i\zeta(u+\delta)}
\biggl( \tanh[u + \delta ] - i\zeta\biggr) .\eqn\nnna$$
Matching these and their
first derivatives at the wall, at $a(u_R) = R,$ gives
$$A_{\zeta}(u ) = \alpha_\zeta B_{ \zeta}(u+ \delta) +
\beta_\zeta B_{-\zeta}(u+ \delta)
\eqn\matching$$
where
$$\eqalign{
\alpha_\zeta &= {2i\zeta - z \over 2i\zeta}e^{-i\zeta \delta} \cr
\beta_\zeta &= {z \over 2 i\zeta} e^{i\zeta(2u_R + \delta)}\cr  }
\eqn\fnc$$
and
$$z = \tanh [u_R]- \tanh [u_R + \delta] = \sqrt{1 - (H_F R)^2} -
\sqrt{1 - (H_T R)^2} .
\eqn\zdefn$$

Note that these become trivial in the limit $\delta \rightarrow 0$, that is,
when the energy difference $(H_F - H_T)\rightarrow 0$. From
the definition of $z$, it appears possible that
$z=0$ is possible even if $H_F \ne H_T$, via taking $R=0$.
However, within the {\it new thin wall} approximation of [\bgt ],
$R \propto \sqrt{H_F^2 - H_T^2}$ with a nonzero constant of proportionality,
and in the usual thin wall approximation [\parke] one has that the
radius obeys
$$R^2 = {S_1^2 \over c_0(H_F^2 - H_T^2) + c_1S_1^2 (H_F^2 + H_T^2)
+ c_2 S_1^4}\eqn\fnd$$
where $S_1$ is the surface tension (related to the integral of the
change in the background field through the wall) and $c_0,c_1,c_2$
are positive constants. Since both $H_F$ and $H_T$ are finite and
the surface tension does not vanish, $R$ is finite.

One actually has to go to another coordinate system in order
to show that there is a `time' where there is only false
vacuum and no bubble [\sasakinew] for initial
conditions.  The consequence here is
that the mode functions must be orthonormalized
once they are continued across the bubble wall, as
they do not correspond to a normalized initial mode functions in this
other `time.'  Thus the mode functions given in eqn.~\matching ~%
are not yet properly orthonormalized.
At the nucleation time
(conventionally taken as the
$\tau=0$ slice in region II), the whole system is rotated to
Lorentzian signature.  The mode functions can then be
orthonormalized in region II,
or since these are subcurvature modes, one can continue
them to regions I and V and orthonormalize them there.
The latter was done in detail in
[\YTSnew, \yssimple] and the former was done in detail in [\jdc].
Substituting the tensor rather than
scalar spatial functions into these results,
in region II the wavefunction is, (again
up to overall factors of $i$),
$$\eqalign{
h^r_{~s}(\xi ,\theta ,\phi ,\eta )=\sum_{jm}\int _0^\infty  &
{d\zeta ~\over \sqrt{ 4\zeta (\zeta^2 +1) \sinh \pi \zeta}}~
\Bigl[ {\bf T}^{Ejm}(\tau +i \pi/2 ,\theta ,\phi ,\eta ;\zeta )
\Bigr] ^r_{~s}\cr
\times \Biggl\{ \hat b_I(\zeta , E, j, m)\cdot \Biggl[
&\Bigl\{ C(+\zeta )~\alpha _{\zeta}+S(+\zeta )~\beta _{-\zeta}\Bigr\} ~
B_\zeta(u)\cr
+ &\Bigl\{ C(+\zeta )~\beta _{\zeta}+S(+\zeta )~\alpha _{-\zeta}\Bigr\} ~
B_{-\zeta}(u)
\Biggr] \cr
+\hat b_V(\zeta , E, j, m)\cdot \Biggl[
&\Bigl\{ C(-\zeta )~\alpha _{-\zeta}+S(-\zeta )~\beta _{\zeta}\Bigr\} ~
B_{-\zeta }(u)\cr
+ &\Bigl\{ C(-\zeta )~\beta _{-\zeta}+S(-\zeta )~\alpha _{\zeta}\Bigr\} ~
B_{\zeta}(u)
\Biggr] ~\Biggr\}  \; . \cr
}\eqn\postwo$$
As a result,
the positive frequency part of the wavefunction inside region I for
subcurvature modes is
$$\eqalign{
h^r_{~s}(\xi ,\theta ,&\phi ,\eta )=\sum_{jm}\int _0^\infty
{d\zeta ~\over \sqrt{ 2\zeta (\zeta^2 +1)}}~
\Bigl[ {\bf T}^{Ejm}(\xi ,\theta ,\phi ,\eta ;\zeta )\Bigr] ^r_{~s}\cr
\times \Biggl\{ \hat b_I(\zeta , E, j, m)\cdot \Biggl[
&\Bigl\{ C(+\zeta )~\alpha _{\zeta}+S(+\zeta )~\beta _{-\zeta}\Bigr\} ~
{e^{+\pi \zeta /2}\over (e^{+\pi \zeta }-e^{-\pi \zeta })^{1/2}}~
T_h(\eta ;+\zeta )\cr
+ &\Bigl\{ C(+\zeta )~\beta _{\zeta}+S(+\zeta )~\alpha _{-\zeta}\Bigr\} ~
{e^{-\pi \zeta /2}\over (e^{+\pi \zeta }-e^{-\pi \zeta })^{1/2}}~
T_h(\eta ;-\zeta )
\Biggr] \cr
+\hat b_V(\zeta , E, j, m)\cdot \Biggl[
&\Bigl\{ C(-\zeta )~\alpha _{-\zeta}+S(-\zeta )~\beta _{\zeta}\Bigr\} ~
{e^{-\pi \zeta /2}\over (e^{+\pi \zeta }-e^{-\pi \zeta })^{1/2}}~
T_h(\eta ;-\zeta )\cr
+ &\Bigl\{ C(-\zeta )~\beta _{-\zeta}+S(-\zeta )~\alpha _{\zeta}\Bigr\} ~
{e^{+\pi \zeta /2}\over (e^{+\pi \zeta }-e^{-\pi \zeta })^{1/2}}~
T_h(\eta ;+\zeta )
\Biggr] ~\Biggr\} \cr
}\eqn\posfreq$$
where
$$\eqalign{D_1(\zeta) &= {1 \over 2} \left[ |\alpha_\zeta|^2 + |\beta_\zeta|^2
+
1 \right] = D_1(-\zeta) = 1 + {z^2 \over 4 \zeta^2},\cr
D_2(\zeta) &= \alpha_\zeta \beta_{\zeta} =
{-z \over 4 \zeta^2}(2i\zeta - z)e^{2 i \zeta u_R},\cr }
\eqn\done$$
and
$$\eqalign{ C(\zeta )&=\sqrt{ D_1\over D_1^2-\vert D_2\vert ^2}~
\sqrt{{1\over 2}\left( 1+\sqrt{1-{\vert D_2\vert ^2\over D_1^2}}\right) }=
\sqrt{{1\over 2}\left( 1+\sqrt{4\zeta ^2\over z^2+4\zeta^2 }\right) }\cr
S(\zeta )&={-D_2\over \vert D_2\vert }~\sqrt{ D_1\over D_1^2-\vert D_2\vert
^2}~
\sqrt{{1\over 2}\left( 1-\sqrt{1-{\vert D_2\vert ^2\over D_1^2}}\right) }=
{-D_2\over \vert D_2\vert }~
\sqrt{{1\over 2}\left( 1-\sqrt{4\zeta ^2\over z^2+4\zeta^2 }\right) }\cr
}\eqn\zzb$$
The identity $D_1^1 - \vert D_2 \vert^2 = D_1$ was used in the above
and can be checked by substituting into the definitions.

For scalars,
an analogue of the vacuum supercurvature mode in the
presence of a bubble, with $\zeta = -i$,
remains normalizable when matched across the wall.\refmark{\yssimple}
This is true for the case here as well, but as
this mode is $\eta$ independent, it does not contribute to the
CMB anisotropy below.  Consequently we do not quote its value here.

\chapter{Implications for the CMB Anisotropy}

Gravitational waves provide a time dependent background
for the photons in the CMB [\sachs] (for a review see
[\cmbrev]).
The contribution of gravitational waves to the CMB anisotropy
is given by the integral in the Sachs--Wolfe formula [\sachs]:
$$\eqalign{
{\delta T_{gw}\over T(\Omega)}&={1\over 2}
\int _{\eta _e}^{\eta _0}d\eta ~\hat r^a\hat r^b
\tilde{h}_{ab,\eta }(\xi =\eta _0-\eta ,\eta )\cr
&={1\over 2}\int _{\eta _e}^{\eta _0}
d\eta ~{~\partial \over \partial \eta }
\tilde{h}_{\hat \xi \hat \xi }(\xi =\eta _0-\eta,\eta )\cr }
\eqn\swf$$
The metric $\tilde{h}$ is defined with the conformal factor scaled out,
hence $\tilde{h}_{\mu \nu} = h_{\mu \nu}/a^2(\eta)$ in our notation.
Here we have chosen the path to be parameterized by conformal
time, where $\eta_0$ is the observing time and $\eta_e$ is the
last scattering time for the photon.  The surface term vanishes
for the tensor contribution and hence is omitted.
The radial-radial component of the mode function has spatial
dependence proportional to $ F_j(\xz),$ as seen in eqn.~\bad .

The Bunch-Davies positive frequency part of the
temperature contrast operator is, using eqn.~\posfreq ~
from the last section,
$$\eqalign{
{\delta T_{gw}^{(+)}(\Omega )\over T_{CMB}}={1\over 2}
\sum _{jm}Y_{jm}&(\Omega )\int _0^\infty d\zeta ~
{1\over \sqrt{2\zeta (\zeta ^2+1)}}
\int _{\eta _e}^{\eta _0}d\eta ~F_j(\xi =\eta _0-\eta ;\zeta )\cr
\times \Biggl\{ \hat b_I(\zeta , E, j, m)\cdot \Biggl[
&\Bigl\{ C(+\zeta )~\alpha_{\zeta }+S(+\zeta )~\beta_{-\zeta }\Bigr\} ~
{e^{+\pi \zeta /2}\over (e^{+\pi \zeta }-e^{-\pi \zeta })^{1/2}}~
\dot T_h(\eta ;+\zeta )\cr
+ &\Bigl\{ C(+\zeta )~\beta_{\zeta}+S(+\zeta )~\alpha_{-\zeta }\Bigr\} ~
{e^{-\pi \zeta /2}\over (e^{+\pi \zeta }-e^{-\pi \zeta })^{1/2}}~
\dot T_h(\eta ;-\zeta )
\Biggr] \cr
+\hat b_V(\zeta , E, j, m)\cdot \Biggl[
&\Bigl\{ C(-\zeta )~\alpha_{-\zeta}+S(-\zeta )~\beta_{\zeta }\Bigr\} ~
{e^{-\pi \zeta /2}\over (e^{+\pi \zeta }-e^{-\pi \zeta })^{1/2}}~
\dot T_h(\eta ;-\zeta )\cr
+
&\Bigl\{ C(-\zeta )~\beta_{-\zeta }+S(-\zeta )~\alpha_{\zeta }\Bigr\} ~
{e^{+\pi \zeta /2}\over (e^{+\pi \zeta }-e^{-\pi \zeta })^{1/2}}~
\dot T_h(\eta ;+\zeta )
\Biggr] ~\Biggr\} \cr
}
\eqn\deltat$$
where the dots indicate derivative with respect to conformal time.
The Bunch-Davies negative frequency component is
$$ {\delta T_{gw}^{(-)}(\Omega )\over T_{CMB}}=
\left[ {\delta T_{gw}^{(+)}(\Omega )\over T_{CMB}}\right] ^\dagger .
\eqn\ssa$$
It is customary to expand the CMB anisotropy in terms of multipoles
according to
$$ {\delta T_{gw}(\Omega )\over T_{CMB}}=
\sum _{lm}a_{lm}~Y_{lm}(\Omega ).\eqn\ssb$$
The statistical average of the ensemble of classical gravity waves
is found by
taking the corresponding quantum average, so that
the two-point correlation is
$$\eqalign{c_l&=\langle \vert a_{lm}\vert ^2\rangle =
{1\over 8}\int _0^\infty {d\zeta \over \zeta (\zeta ^2+1)}
\int _{\eta_e}^{\eta_0}d\eta _1 \int _{\eta_e}^{\eta_0}d\eta _2~
F_l(\xi =\eta _0-\eta _1;\zeta )~ F_l(\xi =\eta _0-\eta _2;\zeta )~\cr
&\times \left\{ \coth [\pi \zeta ] \Re \left[
\dot T_h(\eta_1 ;\zeta )
\dot T_h(\eta_2 ;\zeta )^* \right ]  
+ {1 \over D_1 \sinh [\pi \zeta ]}
\Re \left[ \alpha_{\zeta} \beta_{-\zeta} \dot T_h(\eta_1 ;\zeta )
\dot T_h(\eta_2 ;\zeta ) \right]
\right\} \cr }\eqn\boo$$
for all $l$.  In showing this, it is useful to note that
$\vert C(\zeta) \vert^2 + \vert S(\zeta) \vert^2 = 1$ and
the definitions in eqn.~\done . 
Also since $D_1={1\over 2}
[1+\vert \alpha _\zeta \vert ^2+\vert \beta _\zeta \vert ^2]$ 
and $\vert \alpha_\zeta \vert^2 - \vert \beta_\zeta \vert^2 =1,$
$$ \left| {  2\alpha _\zeta \beta _{-\zeta } \over 
1+\vert \alpha _\zeta \vert ^2+\vert \beta _\zeta \vert ^2}
\right| \le +1.  \eqn\eea$$
Consequently
as one probes larger wavenumbers $\zeta$ 
(through larger-$\ell $ multipoles)
the influence of the bubble dynamics quickly becomes negligible
in a uniform manner.
In other words, effects of the wall radius and finite
energy difference rapidly disappear below the curvature scale,
and thus are expected
to be hidden in the cosmic variance, just as in the case for the
scalar fluctuations.

The integrand in the eqn.~\boo ~for the CMB multipole moments is
well behaved for small $\zeta.$
This can be demonstrated with the explicit form of $F_2 (\xz)$
in eqn.~ \badzf.  
[Recall that for gravity waves the $\ell=0,~1$ moments vanish
because the graviton is a spin-two particle.]
One then notes that $F_j$ for higher $j$ are obtained
by taking derivatives, which will not alter the leading power of
$\zeta$ (although the coefficients may change).
As $\zeta \to 0,$ the time dependence factor in the basis functions
becomes $$\part_\eta T_h(\eta, \zeta)] =
\part_\eta [
(i\zeta \sinh [\eta ]+\cosh [\eta ])e^{-i \zeta \eta}
\rightarrow H_T \sinh [\eta ]~( 1 - i \zeta \eta) + O(\zeta^2) \; .
\eqn\vvd$$
The first term in the curly brackets has the form
$$\eqalign{&\coth [\pi \zeta ]~\Re \left[
\dot T_h(\eta_1 ;\zeta )
\dot T_h(\eta_2 ;\zeta )^* \right ] \cr 
&= \left( {1 \over \pi \zeta}\right) 
H_T^2\sinh^2 [\eta ]~(1 + O(\zeta^2))\cr }
\eqn\first$$
since $\Re [1 - i \zeta(\eta_1 - \eta_2)]=1.$
For the second term in the curly brackets, for small $\zeta ,$
$$\eqalign{&{1 \over D_1 \sinh [\pi \zeta ]}
\Re \left[ \alpha_{\zeta} \beta_{-\zeta} \dot T_h(\eta_1 ;\zeta )
\dot T_h(\eta_2 ;\zeta ) \right] \cr 
&=\left[ 1 + {z^2 \over 4 \zeta^2}\right] ^{-1}{1 + O(\zeta^2)\over \pi \zeta}~~
\Re \left[ \alpha_{\zeta} \beta_{-\zeta} \dot T_h(\eta_1 ;\zeta )
\dot T_h(\eta_2 ;\zeta ) \right] .\cr }\eqn\hhd$$
The argument in brackets on the right has the form
$$\eqalign{&\Re \left[
{2i\zeta - z \over 2i\zeta}~e^{-i\zeta (2\delta+ 2 u_r)}~{-z \over 2 i\zeta}~
\bigl[ 1 - i \zeta (\eta_1 + \eta_2)+ O(\zeta^2)\bigr] \right]
 H_T^2\sinh^2 [\eta ]\cr
&= -{z^2 \over 4 \zeta^2}~\bigl[ 1 +O(\zeta^2)\bigr]H_T^2 \sinh^2 [\eta ]\cr }
\eqn\hhe$$
including the small-$\zeta$ expansion of the exponentials.
Thus, after factoring out the $H_T^2 \sinh^2 [\eta ]$ dependence,
the term in curly brackets in eqn.~\boo ~behaves as
$${1 \over \pi \zeta} + {{-z^2 \over 4 \zeta^2} + O(1)  
\over  \pi \zeta (1 + {z^2 \over 4 \zeta^2})}
\sim \zeta \eqn\limitpower$$
for $\zeta$ small.
In the Sachs-Wolfe integral above,
we also have the factor of $\zeta^{-1}$ in the measure.
Since $N_j \sim \zeta^{-2}$
one has (using eqn.~\badzf ~ in conjunction with eqn.~\xzh ~)
$F_j \sim {\rm constant}.$
This implies that the integrand approaches a constant as $\zeta \to 0,$
rendering the integral infrared convergent.

As $(H_F-H_T)$ becomes small, both $\beta_{-\zeta}$ and 
$z$ approach zero for fixed $\zeta$, and 
the second term in curly brackets approaches zero.
For vanishing $z$, as is found in the vacuum,
eqn.~\limitpower ~approaches $\coth [\pi \zeta ]$,
making the integrand appear to diverge as $\sim \zeta^{-2}.$ 
We do not have an intuitive understanding of this limiting
behavior.   Vanishing $z$ is never the case in the presence of
the bubble since it implies exactly zero energy difference
across the bubble wall.

In order to calculate the $a_{\ell m}$, one needs the time dependence
of the wavefunctions from the inflationary period, through radiation
domination and into the current phase of matter domination.
Matching conditions have been found by [\allen] and
are calculated analogously to the
flat [\gwa-\gwz] and closed [\baclosed]
universe cases.  Time dependence has also been
considered in [\deg] but seemingly for a different initial vacuum.

\chapter{Discussion}

We have determined the initial condition for the
graviton field in an open universe originating from a
bubble inflation model and calculated the contribution
from gravitational waves to the CMB anisotropy.
The total observed CMB anisotropy for a given multipole is obtained
by combining in quadrature the gravity wave contribution calculated here 
with the
scalar field contributions for the particular model.
The effects of the bubble wall for the tensors, just as for the
scalars, seems to be confined mostly to very large scales,
corresponding to $\zeta$ small.  It appears that
the effects of curvature provide a much larger effect for
the tensors than the effects of the wall.\refmark{\wh}

\noindent
{\bf Acknowledgements:} We would like to thank R. Brandenberger,
P. Ferreira, L. Ford, A. Guth, A.  Liddle, A. Linde, B. Ratra,
K. Schleich, N. Turok, and A. Vilenkin
for useful discussions, and especially B. Allen and R.
Caldwell for useful discussions and for sharing their
prior unpublished manuscript\refmark{\allen } with us.
We thank R. Caldwell for comments on the draft.
JDC is grateful in particular to M. White for numerous discussions
and is supported by an ONR grant as a Mary Ingraham Bunting Institute
Science Scholar. JDC also thanks the University of British
Columbia, the Harvard-Smithsonian Center for Astrophysics, and
the Center for Particle Astrophysics, the Physics Department, and
LBNL at Berkeley for hospitality in the course of this work.
MB was supported by the David and Lucille Packard Foundation
and by National Science Foundation grant PHY 9309888.

\noindent
{\bf Noted Added:} After this work was completed we learned that
Allen and Caldwell\refmark{\allentwo } and Sasaki et al.\refmark{\sasakifive }
have reached similar conclusions with respect to the influence of nonvanishing
bubble size.

\refout

\APPENDIX{A}{A---Relation Between Flat and Hyperbolic Coordinates}

For the flat coordinates the line element is (setting $H=1$)
$$ ds^2=-dt_f^2+e^{2t_f}\cdot \Bigl[ dr_f^2+r_f^2~d\Omega _{(2)}^2\Bigr]
\eqn\othmet $$
where $-\infty <t_f<+\infty ,$ or in terms of flat
conformal time $\eta _f=-e^{-t_f}$
$$ ds^2={1\over \eta _f^2}\cdot \Bigl[
-d\eta _f^2+dr_f^2+r_f^2~d\Omega _{(2)}^2\Bigr] .
\eqn\cmetr $$

To relate the flat and hyperbolic coordinates one can
embed de Sitter space in (4+1)-dimensional
Minkowski space (see ref. [\bgt], section 5 or ref. [\hawking]).
The Minkowski coordinates are
$(\bar w, \bar u,\bar x,\bar y,\bar z) = (\bar w,\bar u,\bar r)$ and
de Sitter space is defined by
$$\bar r^2 +\bar u^2 - \bar w^2 = 1.\eqn\ccca$$
The embedding of the open hyperbolic coordinates is
$$\eqalign{
\bar w&=\sinh [t_h]\cosh [\xi ],\cr
\bar u&=\cosh [t_h],\cr
\bar r&=\sinh [t_h]\sinh [\xi ].\cr
}\eqn\ttza$$
Generally, $\bar r >0$, and as $0<t_h< \infty$ and
$0<\xi < \infty$, so that one sees that region I hyperbolic
coordinates cover the range $0 \le {\bar w} \le \infty \, , \;
1 \le {\bar u}$.
The flat coordinates are embedded in (4+1)-dimensional
Minkowski space according to
$$\eqalign{
t_f={\rm ln}[ \bar w+ \bar u], \quad
\eta_f = {-1 \over \bar w+ \bar u}, \quad
r_f={ \bar{r}\over \bar w+ \bar u}.}\eqn\ttza$$
These cover $\bar w, \bar u > 0,$ a larger region than the hyperbolic open
coordinates.  As
$\sinh [t_h]=-1/\sinh [\eta ],$
$\cosh [t_h]=-\coth [\eta ],$
and
$\coth [t_h]=\cosh [\eta ]$,
the relation between the coordinate systems is
$$ \eqalign{
r_f&={\sinh [\xi ]\over \cosh [\xi ]+\cosh [\eta ]},\cr
\eta _f&={\sinh [\eta ]\over \cosh [\eta ]+\cosh [\xi ]}.\cr
}\eqn\flatre$$
The flat coordinates do not cover all of de Sitter space but only
half of maximally extended de Sitter space, as indicated in the
conformal diagram in Fig.~3 in the text.

The null coordinates in the flat chart are
$$\eqalign{
u_f&=\eta_f +r_f,\cr
v_f&=\eta_f -r_f \; . \cr }\eqn\litu$$
One has $v_f<0$ and $\vert u_f \vert<\vert v_f \vert .$ We
also define the region I hyperbolic null coordinates
$$\eqalign{
U&=\eta +\xi ,\cr
V&=\eta -\xi \cr }\eqn\bigU$$
where $V<0$ and $\vert U \vert < \vert V \vert$.
Using the above relations between $(\eta,\xi)$ and
$(\eta_f, r_f)$ one can show that the
two sets of null coordinates are
related according to
$$\eqalign{
U&=\ln \left[ {1+u_f\over 1-u_f} \right] ,\quad
V=\ln \left[ {1+v_f\over 1-v_f} \right] \cr }.\eqn\Udef$$
Conversely,
$u_f = \tanh {U\over 2} $ and $
v_f = \tanh {V\over 2}\; . $
Thus we see that the hyperbolic coordinates cover only the
region $\vert u_f \vert \, , \; \vert v_f \vert \le 1 \, , \;
\vert u_f \vert \le \vert v_f \vert$ in terms of the flat coordinates.

In order to see the continuation into region V, consider the transformation
$u \rightarrow - u$ in the embedding Minkowski space.
This can be accomplished by taking $t_h = -t_V - i\pi$.
In terms of conformal time in region V, this is
$$\eqalign{
\eta_V &= \ln \Bigl[ \tanh (-t_h/2 + i \pi/2)\Bigr]
= \ln(-1) + \ln \Bigl[ \tanh (t_h/2 - i \pi/2)\Bigr] \cr
&= \ln(-1) -\eta_h \cr }\eqn\wsc$$
Thus we have
$$\eqalign{
\tilde{U} = \eta_V + r_V = \ln(-1) - V \cr
\tilde{V} = \eta_V - r_V = \ln(-1) - U \cr }\eqn\wsb$$
where $\tilde{U},\tilde{V}$ are the hyperbolic open coordinates in
region V.  The definition of $\ln(-1)$ requires a choice of
analytic continuation, which has been identified
in terms of $u_f, v_f,$ so converting to these coordinates, one has
$$\eqalign{
u_f = \tanh (U/2) = \tanh(-\tilde{V}/2 - \pm i\pi /2) =
- \coth(\tilde{V}/2) = -\tilde{v}_{f}^{-1} \cr
v_f = \tanh (V/2) = \tanh(-\tilde{U}/2 - \pm i\pi /2) =
-\coth(\tilde{U}/2) = -\tilde{u}_{f}^{-1}
\cr }\eqn\wsa $$
as given in the text.
This transformation relates the flat coordinates $(u_f,v_f)$ covering the
upper left triangular wedge of de Sitter space to
the coordinates $(\tilde u_f,\tilde v_f)$ covering the upper right triangular
wedge, as indicated in Fig.~5.

\APPENDIX{B}{B---Inner Product in Region II}

For this appendix, $H=1$.
To continue into region II,
$e^\eta \rightarrow -i e^{-u},$
$\xi \rightarrow \tau + i \pi/2,$
and the metric becomes
$$ds^2 = a^2(u)~\Bigl[ du^2 - d \tau^2 + \cosh^2 [\tau ]~d\Omega_{(2)}^2\Bigr]
\eqn\met$$
where $a(u) = 1/\cosh [u].$
The Wronskian, with respect to $-i\nabla_\tau$ on the symmetric
perturbations of the metric, gives for an
inner product
$$ - \int _\Sigma {du~d \Omega \cosh^2 [\tau ]\over \cosh^2[u]}~
{\cal U}_A^B(\tau,
\Sigma; \zeta,j,m )~
(i\overleftrightarrow \part_\tau )~
{\cal U}_B^{A*}(\tau,\Sigma; \zeta^\prime,j^\prime,m^\prime )
\eqn\inner$$
The Bunch-Davies vacuum
mode expansion [ see eqn.~\gfa ~] in region I
$$\eqalign{
\sum_{Pjm}\int _0^\infty
&{d\zeta ~\over \sqrt{ 2\zeta (\zeta^2 +1)}}~ 
\Bigl[ {\bf T}^{Ejm}(\xi ,\theta ,\phi ,\eta ;\zeta )\Bigr] ^r_{~s}
\biggl[ {e^{+\pi \zeta /2}\over (e^{+\pi \zeta }-e^{-\pi \zeta })^{1/2}}~~
T_h(\eta ; \zeta)~\hat b_I(\zeta ,P,j,m) \cr
&~~~+{e^{-\pi \zeta /2}\over (e^{+\pi \zeta }-e^{-\pi \zeta })^{1/2}}~~
T_h(\eta; -\zeta) ~\hat b_V%
(\zeta ,P,j,m)\biggr]
+h.c.  \cr
}\eqn\rone$$
becomes in region II
$$\eqalign{&\sum _{PJm}~~\int _0^\infty d\zeta ~ 
\biggl[ {e^{+\pi \zeta /2} \, n(\zeta)
\over (e^{+\pi \zeta }-e^{-\pi \zeta })^{1/2}}~~
T_h\bigl( \ln(-i e^{-u}); \zeta \bigr) ~
T_A^B(\tau + i {\pi\over 2},\theta,\phi ; \zeta,j,m)
{}~\hat b_I(\zeta ,P,J,m)\cr
&~~~+{e^{-\pi \zeta /2} \, n(\zeta)
\over (e^{+\pi \zeta }-e^{-\pi \zeta })^{1/2}}~~
T_h\bigl( \ln(-i e^{-u}); -\zeta\bigr) ~
 [T_A^B(\tau + i {\pi\over 2},\theta,\phi ; \zeta,j,m)]
{}~\hat b_V%
(\zeta ,P,J,m)\biggr] \cr
&+h.c. \}  \cr
&=\sum _{PJm}~~\int _0^\infty d\zeta ~ \{
{\cal U}_A^B (u,\tau,\theta,\phi;\zeta,j,m)
{}~\hat b_I(\zeta ,P,J,m)
+{\cal U}_A^B (u,\tau,\theta,\phi;-\zeta,j,m)
{}~\hat b_V%
(\zeta ,P,J,m) \cr 
&+h.c. \cr } \eqn\ttwo$$
Note that the complex conjugate is only taken in the $u$ dependence
in region II and both terms have the same $\tau$
dependence corresponding to positive frequency.
The integral over the surface $\Sigma$ in region II is an
integral over the analytic continuation of time $u$
and over $(\theta, \phi)$.  The analytic continuation
of $T_h(\eta)$ multiplies the whole expression and is
independent of $\tau $.  So for calculating the
Wronskian, first consider only the $\tau$ dependence and the
integral over $\theta , \phi .$
The integral over $u$ will be done subsequently.
We have
$$\eqalign{
{\bf \tilde{T}}^{E, jm}(\tau ,\theta ,\varphi ; \zeta )&=~~
F_j(\tau + i \pi/2 ; \zeta  )~~({\bf e}^\tau \otimes {\bf e}^\tau )~~
Y_{jm}(\theta ,\varphi )\cr
&~~+ G_j(\tau + i \pi/2 ; \zeta )~\delta _{\tilde a \tilde b}~
({\bf e}^{\tilde a}\otimes {\bf e}^{\tilde b})~
Y_{jm}(\theta ,\varphi ) \cr
&~~+ H_j(\tau + i \pi/2 ; \zeta )~(
{\bf e}^{\tilde a}\otimes {\bf e}^{\tau }
+{\bf e}^{\tau }\otimes {\bf e}^{\tilde a})~
\tilde \nabla _{\tilde a}Y_{jm}(\theta ,\varphi ) \cr
&~~+I_j(\tau + i \pi/2 ; \zeta )~
({\bf e}^{\tilde a}\otimes {\bf e}^{\tilde b})~
\tilde \nabla _{\tilde a}\tilde \nabla _{\tilde b}
Y_{jm}(\theta ,\varphi )\cr }
\eqn\tens$$
Thus
$$\eqalign{
T^\tau_\tau &= - F_j(\tau + i \pi/2 ; \zeta  )~ Y_{jm} \cr
T^\tau_{\tilde{a}} & = -  H_j(\tau + i \pi/2 ; \zeta  )~
\tilde{\nabla}_{\tilde{a}} Y_{jm} \cr
T^{\tilde{a}}_\tau & =
\cosh^{-2}[\tau ]~H_j(\tau + i \pi/2 ; \zeta  )~
\tilde{\nabla}^{\tilde{a}} Y_{jm} \cr
T^{\tilde{a}}_{b} & = \cosh^{-2}[\tau ]~
(I_j (\tau + i \pi/2 ; \zeta  )~
\tilde{\nabla}^{\tilde{a}}\tilde{\nabla}_{\tilde{b}} Y_{jm}
+ \delta^{\tilde{a}}_{\tilde{b}} G_j (\tau + i \pi/2 ; \zeta  )~Y_{jm})
}\eqn\comps$$

As we have $ \nabla^2 Y_{jm} = -j(j+1) Y_{jm} $ and from earlier
$$\eqalign{
\int d\Omega \Bigl( \nabla _{\tilde a}\nabla _{\tilde b}Y_{jm}\Bigr) ^*~
\nabla ^{\tilde a}\nabla ^{\tilde b}Y_{j^\prime m^\prime}
&=-\int d\Omega \Bigl( \nabla ^{\tilde a}\nabla _{\tilde a}\nabla _{\tilde b}
Y_{jm}\Bigr) ^*~ \Bigl( \nabla ^{\tilde b}Y_{j^\prime m^\prime}\Bigr) \cr
&=-\int d\Omega \Bigl( \nabla _{\tilde a}\nabla _{\tilde b}\nabla ^{\tilde a}
Y_{jm}\Bigr) ^*~\Bigl(  \nabla ^{\tilde b}Y_{j^\prime m^\prime}\Bigr) \cr
&=-\int d\Omega \Bigl( \nabla _{\tilde b}\Bigl\{ \nabla ^2 +{1\over 2}R\Bigr\}
Y_{jm}\Bigr) ^*~\Bigl(  \nabla ^{\tilde b}Y_{j^\prime m^\prime}\Bigr) \cr
&=~~[~j(j+1)~]\cdot [~j(j+1)-1~]
 \delta_{j j^\prime} \delta_{m m^\prime} \; ,\cr
}\eqn\iny$$
we obtain immediately that
\def\lrt{\overleftrightarrow \part_\tau}
$$\eqalign{
 \int d \Omega T^A_B \lrt T_A^{B*} & =
\{F_j \lrt F_j^* -2 j(j+1)\cosh^{-2}[\tau ]H_j \lrt H_j^* \cr
&~~+ \cosh^{-4}[\tau ]( I_j \lrt I_j^* (j(j+1))(j(j+1) - 1)
+ 2 G_j \lrt G_j^*  \cr
&~~-j(j+1)(G_j \lrt I_j^* + I_j \lrt G_j^*)) \}
\delta_{j j^\prime} \delta_{m m^\prime} \cr
&=  \{{3\over 2} F_j \lrt F_j^* - 2j(j+1) \cosh^{-2}[\tau ]
H_j \lrt H_j^* \cr
&~~+{1\over 2} \cosh^{-4}[\tau ]J^2(J^2-2) I_j \lrt I_j^* \}
\delta_{j j^\prime} \delta_{m m^\prime}.\cr }
\eqn\wron$$
The sign in front of $H_j$ is due to the negative signature of $\tau$.
There is a factor of two in front of both $H_j$ and $G_j$.  For
$H_j$ there is both a term $T_{1a}$ as well as a term $T_{a1}$, and
for $G_j$ there is a contribution from both $G_{2}$ and $G_{3}$.
The minus sign in front of the last term is due to the negative sign
relating $\nabla^2$ and $J^2$.

After some algebra, one finds that
$$\int d \Omega T^A_B \lrt T_A^{B*}  =
 {2 \zeta^2(1 + \zeta^2) \cosh^4[\tau ]\over j(j+1)(j+2)(j-1)}
F_j \lrt F_j^*\delta_{j j^\prime} \delta_{m m^\prime}
\eqn\wrt$$
After some work one can show that
$$F_j \lrt F_j^* = N_j(\zeta)^2 { i \sinh [\pi \zeta ]
\over \cosh^6 [\tau ]}
\zeta^3 (1 + \zeta^2) (4 + \zeta^2) \cdots (j^2 + \zeta^2).
\eqn\fwron$$
Putting this all together and substituting for $N_j(\zeta)^2,$ we get
$$ \int d \Omega ~T^A_B \lrt T_A^{B*} =
{2 i \zeta  \sinh [\pi \zeta ]\over \pi \cosh^2 [\tau ]}
\delta_{j j^\prime} \delta_{m m^\prime}
\eqn\omint$$
The integral over $u$ remains, with mode functions
$$T_h(\ln(-ie^{-u}); \zeta)
{n(\zeta)e ^{\pi \zeta \over 2} \over \sqrt{e^{+\pi \zeta} - e^{- \pi \zeta})}}
= (i \zeta - \tanh [u])~e^{i \zeta u}
{\cosh [u] \; n(\zeta)\over \sqrt{(e^{\pi \zeta} - e^{- \pi \zeta})}},
\eqn\umode$$
so the last integral, including the rest of the measure, becomes
$$\eqalign{
&\int du \, {\cosh^2 [\tau ]\over \cosh^2 [u]}
{\cosh [u]\bigl(i \zeta -\tanh [u]\bigr) n(\zeta)\over \sqrt{2\sinh [\pi \zeta
]}}
{\cosh [u]\bigl(-i \zeta^\prime -\tanh [u]\bigr) n(\zeta^\prime)\over
\sqrt{2\sinh [\pi \zeta^\prime ]}} e^{i (\zeta-\zeta^\prime) u} \cr
&= {2 \pi \cosh^2 [\tau ]
\over 4 \zeta \sinh [\pi \zeta ]}.}\eqn\lastin$$
As a result, we have
$$- \int _\Sigma \, du \,
\cosh^{-2}[u]\cosh^2 [\tau ]{\cal U}_A^B(u,\Sigma; \zeta,j,m )~
(i\overleftrightarrow \part_\tau )
{\cal U}_B^{A*}(u,\Sigma; \zeta^\prime,j^\prime,m^\prime )
= \delta (\zeta - \zeta^\prime)
\delta_{j j^\prime} \delta_{m m^\prime} \eqn\total $$
which agrees with eqn.~\fhb .

\end